\documentclass[conference]{IEEEtran}
\IEEEoverridecommandlockouts

\usepackage{cite}

\usepackage{amsmath,amssymb,amsfonts}
\usepackage{algorithmic}
\usepackage{graphicx}
\usepackage{textcomp}
\usepackage{xcolor}
\usepackage{tikz}
\usepackage{pgfplots}
\usetikzlibrary{calc,fit,positioning,backgrounds,shapes.geometric,shadows}
\pgfplotsset{compat=1.18}
\usetikzlibrary{shapes, shadows, positioning, backgrounds, fit}
\usepackage{booktabs}
\usepackage{listings}
\definecolor{lstkw}{rgb}{0.4,0.0,0.6}
\definecolor{lststr}{rgb}{0.0,0.5,0.0}
\definecolor{lstcmt}{rgb}{0.4,0.4,0.4}
\definecolor{lstbg}{rgb}{0.96,0.96,0.96}
\lstdefinestyle{pycode}{%
  language=Python,
  basicstyle=\ttfamily\scriptsize,
  keywordstyle=\color{lstkw}\bfseries,
  stringstyle=\color{lststr},
  commentstyle=\color{lstcmt}\itshape,
  backgroundcolor=\color{lstbg},
  frame=single,
  framerule=0.3pt,
  rulecolor=\color{black!35},
  showstringspaces=false,
  breaklines=true,
  numbers=none,
  xleftmargin=4pt,
  xrightmargin=4pt,
  aboveskip=2pt,
  belowskip=2pt,
}
\def\BibTeX{{\rm B\kern-.05em{\sc i\kern-.025em b}\kern-.08em
    T\kern-.1667em\lower.7ex\hbox{E}\kern-.125emX}}
\usepackage{url}
\usepackage[hidelinks]{hyperref}
\usepackage{balance}

\makeatletter
\newcommand{\linebreakand}{%
  \end{@IEEEauthorhalign}
  \hfill\mbox{}\par
  \mbox{}\hfill\begin{@IEEEauthorhalign}
}
\makeatother

\begin{document}

\title{How Generation Architecture Shapes Code Complexity in Multi-Agent LLM Systems: A Paired Study on HumanEval}

\author{\IEEEauthorblockN{Nazmus Ashrafi}
\IEEEauthorblockA{\textit{Independent Researcher} \\
nazmus.s.ashrafi@gmail.com}
}

\maketitle

\begin{abstract}
Large-language-model code generation has shifted from single-shot prompting to multi-agent orchestrations --- analyst, coder, tester, and debugger pipelines --- and is evaluated almost exclusively on functional correctness. Whether these architectures also affect the structural \emph{complexity} of the code they produce, and which orchestration layers carry the cost, remains largely unexamined: prior work has documented prompt-level effects on code complexity, but the architecture-level question is open. We compare six widely-used multi-agent configurations (\texttt{Basic}, \texttt{AC}, \texttt{ACT}, \texttt{Debugger}, \texttt{AC+Debugger}, \texttt{ACT+Debugger}) under two models from the GPT-4o family across all 164 HumanEval tasks --- $1{,}968$ paired observations --- using the five \textsc{radon} complexity metrics (SLOC, cyclomatic complexity, and Halstead Volume, Difficulty, and Effort). We apply a paired non-parametric statistical pipeline (Friedman omnibus, Wilcoxon signed-rank post-hoc with Holm correction, Kendall's $W$ and matched-pairs rank-biserial effect sizes) in both all-completions and passing-only conditions. The six architectures collapse into two indistinguishable complexity clusters separated by a $50$--$130\%$ gap, the same partition in both models and under both conditions; among the architectural layers, the analyst--coder split inflates complexity, the runtime debugger does not --- and on the analyst--coder background actively deflates it --- and the tester re-inflates it. The heavy cluster's additional complexity buys no pass@1 advantage: the leanest architectures match or beat the heaviest on accuracy. Architectural elaboration in LLM code generation should therefore be justified by measured benefit on the dimensions that matter, not assumed.
\end{abstract}


\section{Introduction}
\label{sec:introduction}

The deployment-grade interfaces for LLM-based code generation are no longer single-shot prompts. The strongest performers on HumanEval~\cite{chen2021humaneval} and adjacent benchmarks now wrap the underlying model in multi-agent orchestrations --- an analyst that drafts a plan, a coder that implements it, a tester that critiques, a debugger that executes and repairs --- each layer adding LLM calls, latency, and operational cost in exchange for, ideally, higher correctness~\cite{ashrafi2025multiagent}.

The dominant evaluation lens for these architectures is functional correctness, typically reported as pass@1 on HumanEval or MBPP. In our prior work~\cite{ashrafi2025multiagent} we compared six widely-used multi-agent configurations --- \texttt{Basic}, \texttt{AC}, \texttt{ACT}, \texttt{Debugger}, \texttt{AC+Debugger}, and \texttt{ACT+Debugger} --- across $19$ LLMs and three outcome axes (accuracy, robustness, and latency), and found that adding planning and critique roles (Analyst, Tester) to the pipeline degrades accuracy and robustness, while a runtime debugger remains a comparatively low-cost, high-value component; \texttt{AC+Debugger} emerged as the considered optimum, with the fuller \texttt{ACT+Debugger} chain adding cost without commensurate gain.

Functional correctness, however, is an incomplete scoreboard. The code an architecture emits is also read, reviewed, debugged, and maintained by humans, and its \emph{structural complexity} carries downstream cost --- in comprehension effort, review time, and defect risk --- that pass@1 does not capture. The role of \emph{prompt-level} interventions in shaping that complexity has begun to receive empirical attention: Della Porta et al.~\cite{dinucci2024unlocking}, working on the DEV-GPT corpus of developer--ChatGPT conversations, showed that different prompt patterns (Zero-shot, Few-shot, Chain-of-Thought, Personas) produce code that differs significantly on size-related sub-measures, with Chain-of-Thought consistently the most concise. Their study, however, manipulates prompt phrasing, not generation \emph{architecture}; whether assembling multi-agent pipelines around the same model produces analogous --- or larger --- complexity shifts is unexamined.

Complexity is also known to drift over time and shift across design hierarchy levels (function, class, module, and system), in line with Lehman's laws of software evolution~\cite{lehman1980laws}. Refactoring at one level often relocates complexity to another rather than eliminating it, so a complete picture requires tracking complexity at multiple granularities. We deliberately scope this study to function-level complexity on HumanEval as a controlled baseline; class-, module-, and repository-level effects are left to future work.

The present study provides the first systematic measurement of this architecture-level effect. We ask whether the same six widely-used multi-agent configurations produce code that differs systematically in structural complexity, which of the three architectural \emph{layers} they bundle --- role decomposition ($R$), testing with bounded iteration ($T$), runtime debugging ($D$) --- drive any such effect, whether the effect replicates across the older-flagship and older-affordable variants of the GPT-4o family, and whether it survives conditioning on correctness. We adopt Della Porta et al.'s~\cite{dinucci2024unlocking} dependent-variable battery (the five \textsc{radon} complexity metrics) and statistical recipe (omnibus + post-hoc + effect-size reporting), adapting the latter to our within-task paired design via Friedman and Wilcoxon signed-rank tests with Holm correction (Section~\ref{sec:methodology}).

\textbf{Findings.} Architecture exerts a highly significant effect on every measured complexity dimension, in both models and under both correctness conditions. The six architectures collapse into two internally indistinguishable groups --- a \emph{lean cluster} (\texttt{Basic}, \texttt{Debugger}, \texttt{AC+Debugger}) and a \emph{heavy cluster} (\texttt{AC}, \texttt{ACT}, \texttt{ACT+Debugger}) separated by a $50$--$130\%$ complexity gap. The layer evidence is non-additive: the analyst--coder split inflates complexity, the runtime debugger does \emph{not} (and on the analyst--coder background actively deflates it), and the tester re-inflates it. The pattern reproduces across both models and within the passing-only subset, and the heavy cluster's additional complexity yields no pass@1 advantage. The result corroborates and extends our prior accuracy/robustness finding~\cite{ashrafi2025multiagent} onto the structural-complexity axis: the costly elaboration is conversational, whether from the Analyst's planning or the Tester's critique; the Debugger, also multi-agent but execution-grounded, is not. It also positions generation architecture as a substantially broader lever on code complexity than prompt phrasing: where Della Porta et al.~\cite{dinucci2024unlocking} found prompt patterns to shift only the LOC-related measures, architecture-level intervention shifts all five complexity metrics.

\textbf{Contributions.}
\begin{itemize}
  \item \textbf{C1.} The first systematic measurement of structural complexity across six widely-used multi-agent LLM code-generation architectures, going beyond the correctness-only evaluation that dominates the field.
  \item \textbf{C2.} A paired-design empirical pipeline (Friedman + Wilcoxon signed-rank + Holm correction, with Kendall's $W$ and matched-pairs rank-biserial effect sizes) adapted from Della Porta et al.'s~\cite{dinucci2024unlocking} prompt-level study, with the independent variable swapped from prompt pattern to generation architecture.
  \item \textbf{C3.} Cross-model evidence within the GPT-4o family that the architectural complexity effect is robust across the older flagship (\texttt{gpt-4o-2024-08-06}) and its older affordable sibling (\texttt{gpt-4o-mini-2024-07-18}) --- the cost--capability tradeoff most budget-conscious deployments actually face.
  \item \textbf{C4.} A clean, replicating two-cluster finding with directly practitioner-actionable guidance: the leanest architectures (\texttt{Debugger}, \texttt{AC+Debugger}) match or beat heavier ones on pass@1 while producing markedly simpler code, so architectural elaboration must be justified by measured benefit, not assumed.
\end{itemize}

\textbf{Paper organisation.} Section~\ref{sec:related-work} situates the study against multi-agent code-generation, runtime-debugging, and prompt-pattern complexity literature. Section~\ref{sec:methodology} details the design, the layered architectural framework, and the statistical pipeline. Section~\ref{sec:results} reports the results in research-question order. Section~\ref{sec:discussion} interprets the findings, draws practitioner implications, and discusses threats to validity. Section~\ref{sec:conclusion} concludes and outlines future work.

\section{Related Work}
\label{sec:related-work}

Our work intersects three established lines of research: (i) \emph{multi-agent LLM code generation}, which establishes the architectural population we measure; (ii) \emph{runtime debugging and execution feedback}, which motivates the dialogic-versus-execution-grounded distinction our layered framework isolates; and (iii) the \emph{structural complexity of LLM-generated code}, from which we inherit the dependent-variable battery and statistical recipe.

\subsection{Multi-Agent LLM Code Generation}

LLM code generation has moved beyond single-shot prompting toward multi-agent orchestrations in which specialised LLM-backed roles cooperate to produce, critique, and refine code. MetaGPT~\cite{hong2024metagpt} structures the agent pool as a software-engineering team --- product manager, architect, engineer, tester --- communicating through standardised artefacts; ChatDev~\cite{qian2024chatdev} similarly models a waterfall-like collaboration among role-playing agents. Self-Refine~\cite{madaan2023selfrefine} and Reflexion~\cite{shinn2023reflexion} realise the critique role via single-agent self-reflection rather than role decomposition. MapCoder~\cite{mapcoder2024} adds a retrieval stage to a planning--coding--debugging chain, closes the loop with the problem's sample I/O, and reports strong HumanEval pass@1, with ablations identifying the debugging agent as the single largest contributor to accuracy; AlphaCodium~\cite{alphacodium2024} frames the same shift as a move from \emph{prompt engineering} to \emph{flow engineering} --- a multi-stage code-oriented pipeline that lifts GPT-4 pass@5 on competitive-programming problems from $19\%$ to $44\%$. Across this line of work, two design choices recur: a decomposition of generation into planning, coding, and review roles, and the addition of a feedback loop --- either static (a tester's verdict) or dynamic (runtime execution). Section~\ref{subsec:iv} (Table~\ref{tab:rtd_mapping}) maps these and other systems onto the corresponding architectural layers --- role decomposition ($R$), critique-driven feedback ($T$), and execution-grounded repair ($D$) --- that we vary as our independent variable.

Most directly relevant to this study, our prior work~\cite{ashrafi2025multiagent} evaluated six widely-used configurations of these patterns --- \texttt{Basic}, \texttt{AC}, \texttt{ACT}, \texttt{Debugger}, \texttt{AC+Debugger}, and \texttt{ACT+Debugger} --- on HumanEval and HumanEval+~\cite{liu2023humanevalplus} across $19$ LLMs, measuring functional accuracy (pass@1), robustness (the accuracy drop between the two benchmarks), and latency (end-to-end execution time). That study found that adding agentic roles (moving from a two-agent analyst--coder pair to a three-agent analyst--coder--tester pipeline) generally degrades accuracy and robustness, while a runtime debugger is a comparatively low-cost, high-value component; \texttt{AC+Debugger} emerged as the considered optimum, and the fuller \texttt{ACT+Debugger} chain showed the largest robustness drop, attributed to the ``compounded complexity of multi-agent collaboration and iterative feedback loops.'' That work, however, measured only behavioural outcomes (accuracy, robustness, latency); whether the same architectural layers also shift the \emph{structural complexity} of the code produced --- the question of this paper --- it did not address.

A concurrent line of work characterises \emph{where} multi-agent code-generation systems fail during execution. Cemri et al.~\cite{cemri2025mast} analyse $1{,}642$ execution traces across seven popular frameworks (ChatDev, MetaGPT, HyperAgent, AppWorld, AG2, Magentic-One, OpenManus) and report failure rates of $41$--$87\%$ on the systems' native benchmarks, with $44\%$ of failures attributable to system-design issues, $32\%$ to inter-agent misalignment, and $24\%$ to inadequate task verification. That work documents what goes wrong during MAS execution; ours measures what the structurally successful code actually looks like --- a complementary empirical lens on the same population of architectures.

\subsection{Runtime Debugging and Execution Feedback}

Beyond the role decomposition surveyed above, a complementary line of work treats the execution behaviour of generated code as a signal for repair. LDB~\cite{llmdebugger2024} decomposes a candidate solution along its control-flow graph and re-executes it block-by-block, querying an LLM to judge each block's correctness against the task description and iteratively refining the output. Self-Refine~\cite{madaan2023selfrefine} and Reflexion~\cite{shinn2023reflexion}, mentioned above, likewise close the loop with execution feedback, but with a single critic role rather than the role decomposition characteristic of multi-agent systems. CodeAct~\cite{codeact2024} examines the converse direction at the action-space level: across $17$ LLMs, switching the agent's action format from text or JSON to executable Python code yields up to $+20\%$ absolute success rate with up to $30\%$ fewer interaction turns, with the gap widening as model capability increases.

The architectural distinction between dialogic role decomposition (additional planning and review roles in a conversational loop) and execution-grounded repair (a debugger that judges and re-prompts the coder, with dynamic feedback rather than further conversational roles) is precisely what our layered $\{R, T, D\}$ framework isolates, and prior work has not characterised which kind of feedback regime carries which kind of cost on the produced code. Independent failure-mode analysis on a $1{,}642$-trace MAS corpus identifies \emph{reasoning--action mismatch} ($13.2\%$ of all observed failures) and \emph{task derailment} ($7.4\%$) as among the highest-prevalence inter-agent failure modes~\cite{cemri2025mast} --- failure signatures of dialogic coordination that an execution-grounded debugger inherently sidesteps.

\subsection{Structural Complexity of LLM-Generated Code}

Beyond the system-level interventions covered in the previous two subsections, a parallel literature studies prompt-level interventions on the same complexity metrics. Della Porta et al.~\cite{dinucci2024unlocking} studied how four prompt patterns (Zero-shot, Few-shot, Chain-of-Thought, Personas) influence the structural complexity of Python code generated by ChatGPT on the DEV-GPT corpus. Using \textsc{radon}-derived metrics (LOC and its sub-measures, Cyclomatic Complexity, Halstead Volume/Difficulty/Effort) with Kruskal--Wallis and Dunn's post-hoc with Holm correction, they found significant differences only for LOC-related measures, with Chain-of-Thought consistently producing the most concise code; cyclomatic complexity and the Halstead measures were non-significant across patterns. A follow-up study by the same group on the same corpus extended the analysis to broader code-quality dimensions --- maintainability, security, and reliability --- and likewise found no significant effects of prompt pattern across $7{,}624$ generated files~\cite{dellaporta2025quality}, indicating that the prompt-level signal on the structural and quality properties of generated code is narrow and inconsistent.

Whereas prompt patterns structure the interaction between a user and a single model, generation architectures structure the interaction among specialised reasoning, generation, testing, and debugging components. The present study extends Della Porta et al.'s~\cite{dinucci2024unlocking} line of inquiry from prompt-level to system-level interventions, preserving their dependent-variable battery and statistical recipe while adapting the latter to a within-task paired design (Table~\ref{tab:bridge}). Concurrent work by Idrisov et al.~\cite{idrisov2025program} compares a single-LLM setup to a fixed four-agent AutoGen-based multi-agent setup on six LeetCode problems, reporting cyclomatic complexity, lines of code, and a maintainability index descriptively; the comparison is binary (single-vs-multi) rather than layer-level, and the $24$-observation sample precludes inferential statistics. To our knowledge, no prior work isolates the layer-level effect of agent architecture --- distinct from prompt pattern and from model choice --- on the structural complexity of LLM-generated code, and this paper provides the first systematic measurement at this level.

\begin{table*}[t]
\centering
\caption{This study extends Della Porta et al.~\cite{dinucci2024unlocking} from prompt-level to system-level interventions on LLM-based code generation.}
\label{tab:bridge}
\begin{tabular}{@{}p{0.18\textwidth}p{0.36\textwidth}p{0.42\textwidth}@{}}
\toprule
Study element        & Della Porta et al.~\cite{dinucci2024unlocking} & This study \\
\midrule
Intervention level   & Prompt-level                                    & System-level \\
Independent variable & Prompt pattern                                  & Generation architecture \\
What varies          & User--model interaction protocol                & Architectural layers ($R$, $T$, $D$) \\
Dataset              & DEV-GPT (real developer--ChatGPT conversations) & HumanEval (controlled benchmark) \\
Dependent variables  & LOC, CC, Halstead V/D/E                         & LOC and its sub-measures, CC, Halstead V/D/E \\
Design               & Independent samples per pattern                 & Paired: same task across all six configurations \\
Statistics           & Kruskal--Wallis + Dunn (Holm)                   & Friedman + Wilcoxon (Holm) \\
Primary question     & Do prompt patterns affect complexity?           & Does architecture affect complexity, and which layers drive it? \\
\bottomrule
\end{tabular}
\end{table*}

\section{Methodology}
\label{sec:methodology}

\subsection{Research Questions and Hypotheses}
\label{subsec:rq}

We address three research questions:
\begin{itemize}
  \item \textbf{RQ1:} Within a fixed underlying LLM, do the six generation architectures produce code that differs significantly in structural complexity (SLOC, CC, Halstead V/D/E)?
  \item \textbf{RQ2:} Does any architectural complexity effect identified under RQ1 replicate across the older-flagship and older-affordable variants of the GPT-4o family?
  \item \textbf{RQ3:} Are any complexity differences identified under RQ1 robust to conditioning on correctness, or are they artefacts of architecture-specific failure-mode behaviour?
\end{itemize}

For each complexity metric $m \in \{$SLOC, CC, $V$, $D$, $E\}$ we test the pair $H_{m,0}$: the six architectures yield equal distributions of $m$ within tasks; $H_{m,A}$: at least one architecture differs. Each pair is tested independently per model under the primary all-completions analysis and re-tested under the secondary passing-only robustness analysis (Section~\ref{subsec:passcond}), yielding $2 \times 5 \times 2 = 20$ omnibus tests --- 10 primary, 10 secondary; all post-hoc comparisons are two-sided. Our prior work~\cite{ashrafi2025multiagent}, which evaluated these same six configurations, found that increasing agentic elaboration --- adding roles to the pipeline --- degrades functional accuracy and robustness, while runtime debugging remains a comparatively low-cost component; that study measured accuracy, robustness, and latency, but not the structural complexity of the generated code. RQ1 asks whether agentic elaboration also carries a structural-complexity cost.

\subsection{Independent Variable: Generation Architecture (Six Configurations)}
\label{subsec:iv}


\begin{figure*}[t]
\centering
\definecolor{hdrR}{RGB}{74,106,138}      
\definecolor{hdrT}{RGB}{74,122,74}       
\definecolor{hdrD}{RGB}{138,90,42}       
\definecolor{ioFill}{RGB}{245,245,245}   
\definecolor{attrFill}{RGB}{248,248,248} 
\definecolor{bdr}{RGB}{85,85,85}         
\definecolor{arwC}{RGB}{68,68,68}        
\definecolor{agBdr}{RGB}{153,153,153}    

\begin{tikzpicture}[
  font=\footnotesize,
  >=stealth,
  agent/.style={
    rectangle, rounded corners=3pt, draw=agBdr, fill=white,
    minimum height=11mm, minimum width=22mm, align=center,
    font=\scriptsize, inner sep=2.5pt},
  substage/.style={
    rectangle, rounded corners=3pt, draw=agBdr, fill=white,
    minimum height=11mm, minimum width=19mm, align=center,
    font=\scriptsize, inner sep=2pt},
  io/.style={
    rectangle, rounded corners=3pt, draw=bdr, fill=ioFill,
    minimum height=8mm, minimum width=130mm, align=center,
    font=\scriptsize, inner sep=2.5pt},
  panel/.style={
    rectangle, rounded corners=3pt, draw=bdr,
    fill=white, inner sep=6pt},
  hdr/.style={
    rectangle, rounded corners=1.5pt, anchor=south west,
    font=\scriptsize\bfseries\color{white},
    inner xsep=4pt, inner ysep=1.5pt},
  attrbox/.style={
    rectangle, rounded corners=3pt, draw=agBdr!70,
    fill=attrFill, text width=38mm, align=left,
    font=\scriptsize, inner sep=4pt},
  flow/.style={->, semithick, draw=arwC},
  loopa/.style={->, semithick, draw=arwC, dashed},
  lbl/.style={font=\tiny, text=arwC, fill=white, inner sep=0.5pt}
]

\node[io] (input) at (0,0) {%
  \textbf{Code Requirements:}\;
  task description $P$,\; visible tests $T_v$};

\node[agent] (analyst) at (-45mm,-22mm) {%
  \textbf{Analyst}\\[0.5pt]
  {\tiny Decompose $P$}\\[-1pt]
  {\tiny into subproblems}};

\node[agent, right=12mm of analyst] (coderR) {%
  \textbf{Coder}\\[0.5pt]
  {\tiny Generate seed}\\[-1pt]
  {\tiny program $A_0$}};

\draw[flow] (analyst) -- node[above, lbl] {Plan} (coderR);

\node[attrbox, right=12mm of coderR] (attrR) {%
  \textbf{Role = Analyst + Coder}\\[1.5pt]
  --\;\underline{Feedback}: none\\[1pt]
  --\;\underline{Iteration}: single-pass\\[1pt]
  --\;\underline{Output}: seed program $A_0$};

\begin{scope}[on background layer]
  \node[panel, fit=(analyst)(coderR)(attrR),
        inner xsep=8pt, inner ysep=6pt] (panelR) {};
\end{scope}
\node[hdr, fill=hdrR] at (panelR.north west)
  {Layer $R$\;---\;Role Decomposition};

\draw[flow] (input.south -| coderR) -- (panelR.north -| coderR);

\node[agent] (tester) at
  ($(panelR.south -| analyst) + (0,-18mm)$) {%
  \textbf{Tester}\\[0.5pt]
  {\tiny LLM-based static}\\[-1pt]
  {\tiny review (no exec.)}};

\node[agent, right=12mm of tester] (coderT) {%
  \textbf{Coder}\\[0.5pt]
  {\tiny Revise code from}\\[-1pt]
  {\tiny tester report}};

\draw[flow] ([yshift=1.5mm]tester.east) --
  node[above, lbl] {Report}
  ([yshift=1.5mm]coderT.west);
\draw[loopa] ([yshift=-1.5mm]coderT.west) --
  node[below, lbl] {Revised code}
  ([yshift=-1.5mm]tester.east);

\node[font=\tiny, text=arwC] (loopTlbl)
  at ($(tester.south)!0.5!(coderT.south) + (0,-2mm)$)
  {$\le 3$ iterations};

\node[attrbox, right=12mm of coderT] (attrT) {%
  \textbf{Role = Tester + Coder}\\[1.5pt]
  --\;\underline{Feedback}: static\\
  \quad (LLM review, no execution)\\[1pt]
  --\;\underline{Iteration}: $\le 3$ rounds\\[1pt]
  --\;\underline{Output}: validated code};

\begin{scope}[on background layer]
  \node[panel, fit=(tester)(coderT)(attrT)(loopTlbl),
        inner xsep=8pt, inner ysep=6pt] (panelT) {};
\end{scope}
\node[hdr, fill=hdrT] at (panelT.north west)
  {Layer $T$\;---\;Testing + Bounded Iteration};

\draw[flow] (panelR.south -| coderR) --
  node[right, lbl, pos=0.45] {$A_0$}
  (panelT.north -| coderR);

\node[substage] (profile) at
  ($(panelT.south -| analyst) + (0,-18mm)$) {%
  \textbf{Profiling}\\[0.5pt]
  {\tiny CFG $\to$ blocks $G_i$}\\[-1pt]
  {\tiny Execute, capture $S_i$}};

\node[substage, right=5mm of profile] (dbg) {%
  \textbf{Debugging}\\[0.5pt]
  {\tiny LLM batch verdicts}\\[-1pt]
  {\tiny $\{v_i, x_i\}$ per block}};

\node[substage, right=5mm of dbg] (regen) {%
  \textbf{Regeneration}\\[0.5pt]
  {\tiny LLM re-generates}\\[-1pt]
  {\tiny from block feedback}};

\draw[flow] (profile) --
  node[above, font=\tiny, text=arwC] {trace} (dbg);
\draw[flow] (dbg) --
  node[above, font=\tiny, text=arwC] {verdicts} (regen);

\node[substage, below=6mm of regen] (execD) {%
  \textbf{Execute}\\[0.5pt]
  {\tiny Run visible tests}};

\draw[flow] (regen) -- (execD);

\draw[loopa, rounded corners=3pt]
  (execD.west) -|
  node[pos=0.2, below, font=\tiny, text=arwC]
    {Fail ($\le\!4\times$)}
  (profile.south);

\node[attrbox, right=8mm of regen] (attrD) {%
  \textbf{Role = Debugger + Coder}\\[1.5pt]
  --\;\underline{Feedback}: dynamic\\
  \quad (block-wise trace +\\
  \quad runtime variables)\\[1pt]
  --\;\underline{Iteration}: $\le 4$ repair loops\\[1pt]
  --\;\underline{Output}: refined $A^*$};

\begin{scope}[on background layer]
  \node[panel, fit=(profile)(dbg)(regen)(execD)(attrD),
        inner xsep=8pt, inner ysep=6pt] (panelD) {};
\end{scope}
\node[hdr, fill=hdrD] at (panelD.north west)
  {Layer $D$\;---\;Runtime Debugging};

\draw[flow] (panelT.south -| coderR) --
  node[right, lbl, pos=0.45] {validated code}
  (panelD.north -| coderR);

\node[io, below=7mm of panelD.south] (output) {%
  \textbf{Final $A^*$}\;
  evaluated with hidden tests $T_h$};

\draw[flow] (panelD.south -| coderR) --
  node[right, lbl, pos=0.45] {$A^*$}
  (output.north -| coderR);

\end{tikzpicture}
\caption{The \texttt{ACT+Debugger} pipeline as the union of three architectural layers:
$R$~(role decomposition: Analyst\,+\,Coder),
$T$~(testing with static LLM-based code review and bounded iteration), and
$D$~(runtime debugging with block-wise execution feedback and repair loop).
Each of the six configurations is a subset of $\{R,T,D\}$ (Table~\ref{tab:configs});
\texttt{Basic}~$= \varnothing$.
Solid arrows show data flow; dashed arrows show iteration loops.
\textit{Notation:} $P$~task description, $T_v$/$T_h$~visible/hidden tests,
$A_0$~seed program, $A^*$~refined output;
$G_i$~control-flow blocks, $S_i$~runtime variable state after block $G_i$,
$\{v_i,x_i\}$~per-block correctness verdict and explanation.}
\label{fig:architecture}
\end{figure*}
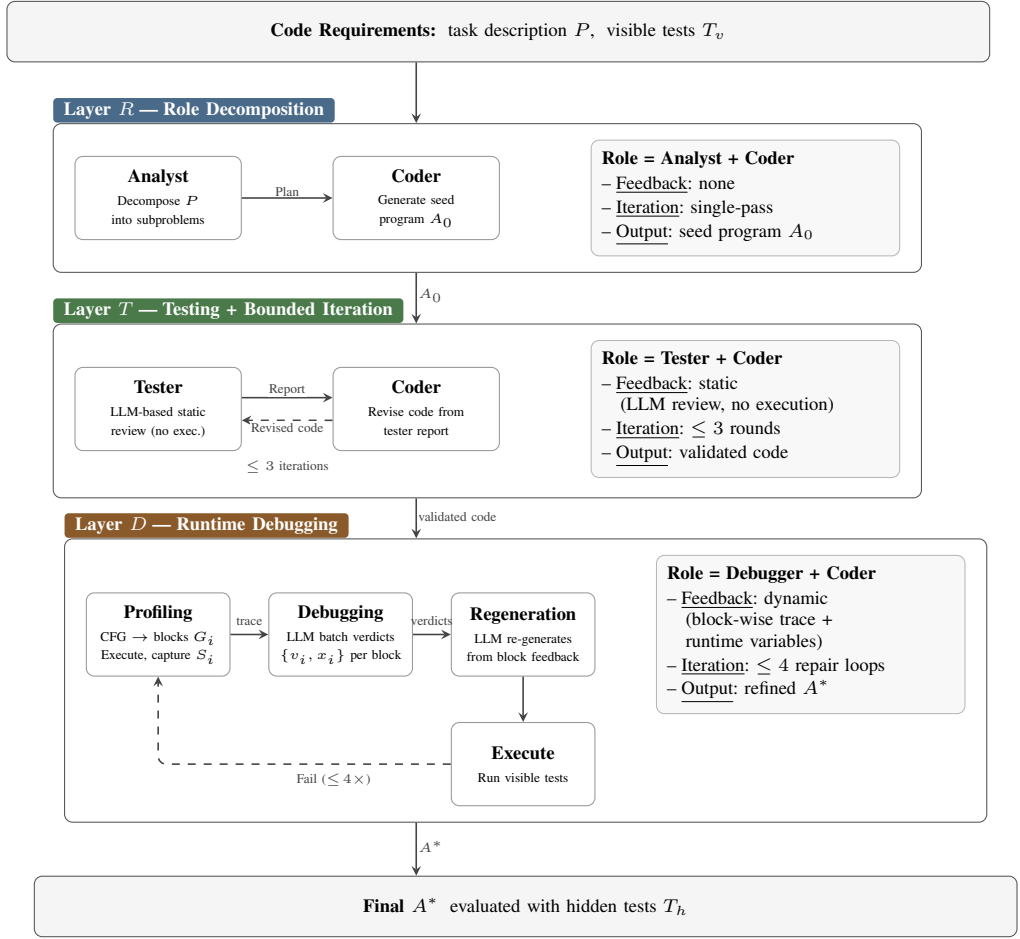

The independent variable in this study is the \textbf{LLM code-generation architecture}, operationalised through the six configurations introduced in our prior work~\cite{ashrafi2025multiagent}. These configurations are not arbitrary engineering variants but combinations of three architectural \textit{layers} (Fig.~\ref{fig:architecture}), each adding one specialised role on top of the always-present Coder together with the feedback regime and iteration regime that role brings (Table~\ref{tab:configs}):
\begin{itemize}
  \item \textbf{$R$ --- Role decomposition.} An Analyst is added that drafts a plan before the Coder generates code; $R$ contributes the planning role.
  \item \textbf{$T$ --- Testing with bounded iteration.} A Tester critiques the Coder's output, returns static feedback, and triggers up to three refinement rounds.
  \item \textbf{$D$ --- Runtime debugging.} A Debugger executes the candidate solution, ingests dynamic execution feedback, and runs a bounded repair loop in which the Coder is re-prompted to regenerate.
\end{itemize}

Throughout, we use \emph{agent} in the operational sense recurring in the multi-agent code-generation literature mapped in Table~\ref{tab:rtd_mapping}: a role-specialized LLM call (or, for the Debugger, a small sub-pipeline) with a fixed system prompt, coordinated with other agents through a workflow graph --- without autonomous goal-setting, free-form tool use, or ReAct-style~\cite{yao2023react} action loops. Figure~\ref{fig:architecture} traces the full \texttt{ACT+Debugger} pipeline end-to-end; the other configurations are obtained by removing one or more layer panels. A task enters at the input strip as a description $P$ paired with visible tests $T_v$. Layer $R$ activates the Analyst, which decomposes $P$ into a plan and hands it to the Coder; the Coder produces the seed program $A_0$ in a single pass. Layer $T$ (when present) routes $A_0$ to the Tester, which performs LLM-based static review without execution, returns a report, and triggers up to three Tester--Coder revision rounds, yielding validated code. The validated code is then executed against $T_v$: if all visible tests pass, the pipeline emits it as $A^*$ and Layer $D$ is bypassed entirely; only on a visible-test failure does Layer $D$ engage (the same gate sits between any upstream output and Layer $D$ in configurations without $T$). Layer $D$ (when present) runs a three-stage repair loop on that code: \textit{Profiling} segments the program along its control-flow graph into basic blocks $G_i$, executes it against $T_v$, and captures runtime variable state $S_i$; \textit{Debugging} queries the LLM in batch for per-block correctness verdicts and explanations $\{v_i, x_i\}$; \textit{Regeneration} prompts the LLM to rewrite the program from the block-level feedback. The repaired program is re-executed against $T_v$; failed runs re-enter Profiling, bounded by a configuration-specific cap (\texttt{Debugger} alone: up to 10 loops, mirroring LDB~\cite{llmdebugger2024}; \texttt{AC+Debugger} and \texttt{ACT+Debugger}: up to 4). The refined output $A^*$ exits at the bottom and is evaluated against the hidden tests $T_h$ to determine pass@1.

All configurations except \texttt{Basic} are therefore multi-agent in implementation: \texttt{Basic} is the single-call Coder, and every other configuration composes one or more of $\{R, T, D\}$ onto that Coder substrate --- including \texttt{Debugger}, whose repair loop re-prompts the Coder for each regeneration. The Coder is the constant across configurations; each layer adds one specialised role around it.
A configuration is the binary presence vector $(R, T, D) \in \{0,1\}^3$. The six configurations populate six of the eight cells of this cube; the two unfilled cells (those with $T=1$ but $R=0$) are not engineering-meaningful, because the testing layer requires a coder whose output it can test. Concretely: \texttt{Basic} is the empty baseline $(0,0,0)$, \texttt{AC} is $R$ only $(1,0,0)$, \texttt{ACT} is $R\!+\!T$ $(1,1,0)$, \texttt{Debugger} is $D$ only $(0,0,1)$, \texttt{AC+Debugger} is $R\!+\!D$ $(1,0,1)$, and \texttt{ACT+Debugger} is $R\!+\!T\!+\!D$ $(1,1,1)$. Treating the six configurations as a layered architectural design space lets us study whether system-level orchestration of LLM code generation produces complexity effects analogous to those Della Porta et al.~\cite{dinucci2024unlocking} observed for prompt-level interventions.

Because each layer bundles a role, a feedback regime, and an iteration regime, the three layers are components, not orthogonal feature axes: adding $D$ always brings dynamic execution feedback and a repair loop together with the debugger role. Pairwise comparisons across the six configurations therefore either toggle a single layer on a fixed background (\textit{Single}), add or remove more than one layer in the same direction (\textit{Compound}), or exchange one layer for another with the third held fixed (\textit{Swap}; e.g., \texttt{AC} vs \texttt{Debugger}). We classify all 15 pairwise comparisons in advance (Section~\ref{subsec:mechanisms}) and restrict causal-attribution claims to single-layer pairs.

\begin{table}[t]
\centering
\footnotesize
\caption{The six configurations as combinations of three architectural layers: $R$ (role decomposition), $T$ (testing + bounded iteration), and $D$ (runtime debugging). The right-hand columns enumerate what each present layer contributes.}
\label{tab:configs}
\begin{tabular}{@{}lllll@{}}
\toprule
Configuration & Layers & Roles & Feedback & Iteration \\
\midrule
\texttt{Basic}        & ---               & ---       & ---          & Single-pass \\
\texttt{AC}           & $R$               & A+C       & ---          & Single-pass \\
\texttt{ACT}          & $R\!+\!T$         & A+C+T     & Static       & $\le 3$ iters. \\
\texttt{Debugger}     & $D$               & D         & Dynamic      & Repair loop \\
\texttt{AC+Debugger}  & $R\!+\!D$         & A+C+D     & Dynamic      & Plan+repair \\
\texttt{ACT+Debugger} & $R\!+\!T\!+\!D$   & A+C+T+D   & Static+Dyn.  & Iter.+repair \\
\bottomrule
\end{tabular}
\\[2pt]
{\scriptsize A = Analyst, C = Coder, T = Tester, D = Debugger.}
\end{table}

The other half of the taxonomic argument is that the three layers are themselves the recurring design choices in this literature, not an arbitrary trio. Table~\ref{tab:rtd_mapping} maps the $(R, T, D)$ functional roles onto representative multi-agent and ACI-based single-agent code-generation systems. Distinct systems use different names for the same functional role --- MetaGPT's~\cite{hong2024metagpt} \emph{QA Engineer}, ChatDev's~\cite{qian2024chatdev} \emph{Reviewer}, AgentCoder's~\cite{huang2024agentcoder} \emph{Test Designer}, AlphaCodium's~\cite{alphacodium2024} \emph{AI Test Generator}, and SWE-agent's~\cite{sweagent2024} \emph{linter} all instantiate $T$ in different operational forms --- and the recurrence of the three layers across systems built independently by different research groups supports treating $(R, T, D)$ as a taxonomy of the design space's recurring choices rather than a post-hoc classification of our own pipeline.

\begin{table*}[t]
\centering
\footnotesize
\caption{Functional mapping of the $(R, T, D)$ architectural layers onto representative multi-agent (and ACI-based single-agent) code-generation systems. Cell entries are each system's own term for the role; ``---'' indicates the functional layer is absent. The recurrence of role decomposition ($R$), critique-driven feedback ($T$), and execution-grounded repair ($D$) across systems built independently of this study supports treating $(R, T, D)$ as a description of the design space's recurring choices rather than an arbitrary classification.}
\label{tab:rtd_mapping}
\begin{tabular}{@{}llll@{}}
\toprule
System & $R$ (planning role) & $T$ (critique-driven feedback) & $D$ (execution-grounded repair) \\
\midrule
\multicolumn{4}{@{}l}{\textit{Multi-agent code-generation systems}} \\
MetaGPT~\cite{hong2024metagpt}            & Architect + PM             & QA Engineer                  & ---                             \\
ChatDev~\cite{qian2024chatdev}            & CEO + CTO                  & Reviewer                     & ---                             \\
AgentCoder~\cite{huang2024agentcoder}     & ---                        & Test Designer                & Test Executor                   \\
MapCoder~\cite{mapcoder2024}              & Planning Agent             & (plan confidence)            & Debugging Agent                 \\
AlphaCodium~\cite{alphacodium2024}        & Reflection + Ranking       & AI Test Generator$^\dagger$  & Iterate-on-tests                \\
\midrule
\multicolumn{4}{@{}l}{\textit{Single-agent systems ($R/T/D$ as tooling or self-feedback)}} \\
Self-Refine~\cite{madaan2023selfrefine}   & ---                        & Self-critique                & ---                             \\
Reflexion~\cite{shinn2023reflexion}       & ---                        & Verbal RL critique           & partial                         \\
LDB~\cite{llmdebugger2024}                & ---                        & ---                          & Block-trace verdict             \\
SWE-agent~\cite{sweagent2024}$^\ddagger$  & (implicit, ReAct thought)  & Linter on \texttt{edit}      & Reproduce-script-then-fix       \\
\midrule
\textbf{\texttt{ACT+Debugger} (this study)} & \textbf{Analyst}         & \textbf{Tester}              & \textbf{Debugger}               \\
\bottomrule
\end{tabular}
\\[2pt]
{\scriptsize $^\dagger$ AlphaCodium's $T$ generates additional test cases rather than producing verdict messages; same functional role, different operational behaviour.\hspace{1em}$^\ddagger$ SWE-agent is single-agent; $R/T/D$ appear as interface tooling rather than separate agents.}
\end{table*}

\subsection{Models}
We evaluate two models from the GPT-4o family: \textbf{GPT-4o} (\texttt{gpt-4o-2024-08-06}, the previous-generation flagship) and \textbf{GPT-4o-mini} (\texttt{gpt-4o-mini-2024-07-18}, its cost-efficient variant). The pair is deliberately drawn from a single provider's family rather than spanning closed- vs.\ open-source: holding pretraining lineage and alignment recipe roughly fixed isolates the architectural effect from the confounds a cross-provider comparison would introduce (different pretraining corpora, different RLHF, different decoding defaults), while still exercising the capability gradient practitioners trade against cost. The pair also reflects a realistic deployment dichotomy: a previous-generation flagship and its cost-efficient sibling are the two endpoints most budget-conscious teams pick between in production. Cross-provider and open-source generalisation is left to future work and discussed under threats to validity (Section~\ref{sec:discussion}). All decoding uses temperature $= 0$, following the prior work's protocol~\cite{ashrafi2025multiagent}.

\subsection{Benchmark and Dataset}
We use \textbf{HumanEval}~\cite{chen2021humaneval}, comprising 164 hand-written Python programming problems, each consisting of a function signature, a natural-language docstring, and a hidden reference test suite. Della Porta et al.~\cite{dinucci2024unlocking} evaluated prompt patterns on DEV-GPT, a corpus of real-world developer--ChatGPT conversations. Because our independent variable is generation architecture rather than prompt phrasing, controlled and reproducible IV manipulation requires problems whose specification, evaluation criteria, and execution environment are fixed --- HumanEval provides this, DEV-GPT does not. We therefore adopt Della Porta et al.'s~\cite{dinucci2024unlocking} dependent-variable pipeline and statistical recipe while using HumanEval as the substrate. The implication --- that the resulting populations of generated code are not directly comparable across the two studies --- is addressed in Section~\ref{sec:discussion} (Threats to Validity). Each of the 164 tasks is solved under every (model, architecture) combination, producing 12 paired observations per task and a total of $164 \times 2 \times 6 = 1{,}968$ generated solutions.

\subsection{Dependent Variables: Complexity Metrics}
All complexity metrics are computed with the \textsc{radon} library~\cite{radon}, mirroring Della Porta et al.~\cite{dinucci2024unlocking}. Metrics are calculated on the \textbf{model-generated completion only}; the original HumanEval prompt (function signature, docstring, examples) and the hidden test cases are excluded from the analysed code, isolating the agent's contribution from dataset-supplied scaffolding.

\begin{itemize}
  \item \textbf{LOC}, decomposed into \textit{Source Lines of Code} (SLOC), \textit{Multi-line strings} (Multi), \textit{Comments}, and \textit{Blank lines}. Aggregate LOC is reported alongside the four sub-measures, since Della Porta et al.'s~\cite{dinucci2024unlocking} significant LOC effects appeared at the sub-measure level.
  \item \textbf{Cyclomatic Complexity (CC)}~\cite{mccabe1976complexity} --- the number of linearly independent paths through the code.
  \item \textbf{Halstead Volume} ($V = N \cdot \log_2 n$), \textbf{Difficulty} ($D$), and \textbf{Effort} ($E = V \cdot D$)~\cite{halstead1977elements}.
\end{itemize}

For each generated completion we record two data-quality flags \textit{separately}: \textit{parse validity} --- the output parses as Python after fence stripping and surrounding-text removal; and \textit{entry-point presence} --- the parsed AST defines a function matching the task's expected \texttt{entry\_point} name. The two signal distinct failure modes: parse-invalid completions indicate generation breakdown, while entry-point-missing completions indicate an intent-to-name mismatch that may itself be partially recoverable but is treated here as an exclusion for analytical consistency. Per-configuration rates of each failure mode are reported alongside the complexity summaries (Section~\ref{subsec:dataquality}), since configurations that fail more frequently are themselves less reliable and this reliability signal must not be lost in the listwise-deleted statistical analysis.

\subsection{Pass-Conditional Robustness Analysis}
\label{subsec:passcond}

We run two complementary analyses in a primary/secondary configuration; they are not co-equal.

The \textbf{primary analysis (all-completions)} includes every parse-valid generated solution regardless of test outcome. It answers the deployment-realistic question --- what does the code an architecture actually produces look like when run on a fresh task? --- and preserves the maximum block count for the paired Friedman/Wilcoxon stack.

The \textbf{secondary analysis (passing-only)} includes only tasks for which all six architectures, under the same model, produce a solution that passes the reference tests. Its purpose is \textit{failure-mode disambiguation}: a complexity difference observed under all-completions could plausibly be driven by an architecture's failure mode rather than its successful-code behaviour --- for instance, an architecture that fails by emitting verbose, over-elaborated repair attempts would register more complex code on average without that complexity reflecting how it solves the problems it actually solves. Comparing the two analyses identifies which all-completions effects survive among solutions that every architecture got right (robust) and which evaporate (failure-mode artefact). The passing-only result is read as a robustness check on the primary, not as a parallel finding in its own right.

The passing-only filter is strict by construction --- a task is excluded whenever any of the six architectures fails on it --- which preserves the paired-block design Friedman and Wilcoxon require at the cost of discarding tasks where most architectures succeed. Retained $n$ is reported for every test; where it falls below a level supporting reliable inference for a given (model, metric) cell, we report descriptive statistics and omit the inferential test for that cell, flagging the cell explicitly.

\subsection{Statistical Analysis}
\label{subsec:stats}

\subsubsection{Rationale for Non-Parametric Repeated-Measures Methods}
Our design is paired: the same 164 HumanEval tasks are solved under every (model, architecture) combination, so the six per-task complexity values are not independent across architectures. This rules out one-way ANOVA and Kruskal--Wallis, which assume independent groups. The parametric default for six paired conditions, repeated-measures ANOVA, additionally requires approximately normal residuals and sphericity of within-task differences. The dependent variables in this study are either discrete and count-valued (SLOC and its sub-measures, CC) or right-skewed by construction (Halstead $V, D, E$ are unbounded above and grow super-linearly with code size), and prior reports on the same metric family on LLM-generated code document substantial right skew and outliers~\cite{dinucci2024unlocking}. We therefore adopt a non-parametric repeated-measures workflow a priori, using the paired analogues of every test in Della Porta et al.'s~\cite{dinucci2024unlocking} independent-samples pipeline. Holm correction and the conceptual decomposition into omnibus + post-hoc are preserved from their recipe; the switch from Kruskal--Wallis/Dunn to Friedman/Wilcoxon reflects the design difference, not a departure from their analytical spirit.

\subsubsection{Omnibus: Friedman's Test}
For each (model, metric, correctness condition) we test $H_{m,0}$ with Friedman's test~\cite{friedman1937}. Friedman ranks the six architectural values 1--6 within each task block, sums ranks $R_i$ per architecture across blocks, and computes
\[
Q = \frac{12}{n\,k\,(k+1)}\sum_{i=1}^{k} R_i^2 \;-\; 3\,n\,(k+1),
\]
where $k = 6$ is the number of architectures and $n$ is the number of complete task blocks retained after the deletion rule below. Under $H_{m,0}$, $Q$ follows $\chi^{2}_{k-1}$ asymptotically. Significance is declared at $\alpha = 0.05$.

\subsubsection{Post-hoc: Wilcoxon Signed-Rank with Holm Correction}
Where an omnibus rejects, we compute Wilcoxon signed-rank tests~\cite{wilcoxon1945} for all $\binom{6}{2} = 15$ architectural pairs within that (model, metric, correctness) family. To control the family-wise error rate within each post-hoc family of 15, we apply Holm's step-down procedure~\cite{holm1979}, which is uniformly more powerful than Bonferroni while maintaining strict FWER control. A pairwise difference is reported significant only if its Holm-adjusted $p < 0.05$. Holm correction is applied within each post-hoc family, not across the full 20-omnibus grid, mirroring Della Porta et al.'s~\cite{dinucci2024unlocking} reporting convention; we acknowledge under threats to validity that this is the standard but not the only defensible choice (Section~\ref{sec:discussion}).

\subsubsection{Effect Sizes}
Statistical significance does not entail practical importance. For each omnibus we report Kendall's coefficient of concordance, $W = Q / (n\,(k-1))$ --- the paired-design analogue of the Rank $\varepsilon^{2}$ that Della Porta et al.~\cite{dinucci2024unlocking} report for their Kruskal--Wallis omnibus --- interpreted on the conventional non-parametric scale (weak $W < 0.3$, moderate $0.3 \le W < 0.5$, strong $W \ge 0.5$). For each significant pairwise comparison we report the matched-pairs rank-biserial correlation,
\[
r_{rb} = \frac{W^{+} - W^{-}}{W^{+} + W^{-}},
\]
where $W^{+}$ and $W^{-}$ are the sums of positive and negative signed ranks~\cite{wilcoxon1945,kerby2014}, interpreted on the conventional scale (small $\approx 0.1$, moderate $\approx 0.3$, large $\ge 0.5$). Reporting both significance and effect size addresses recurrent criticisms of effect-size-blind significance testing in empirical software engineering.

\subsubsection{Missing Data and Block Construction}
Friedman requires complete blocks. We apply listwise deletion at the task level \textit{within each (model, metric, correctness condition) family}: a task is dropped from that family if any of the six architectures, on that model, failed to produce a parse-valid completion with the expected entry point on that task --- or, additionally for the passing-only analysis, if any of the six failed the reference tests. Deletion is per-family rather than global, so the effective $n$ may differ across (metric, correctness) cells. The retained $n$ is reported for every test, and the per-configuration rate of each exclusion cause (parse-invalid, entry-point-missing, test-failing) is reported alongside as an irreducible generation-reliability signal.

\subsubsection{Software}
All tests are computed in Python using \texttt{scipy.stats.friedmanchisquare} and \texttt{scipy.stats.wilcoxon}; Holm correction uses \texttt{statsmodels.stats.multitest.multipletests}. Kendall's $W$ and matched-pairs rank-biserial correlations are derived directly from test outputs.

\subsection{Layer Isolation in Post-hoc Interpretation}
\label{subsec:mechanisms}

Because the six configurations differ in which of the three architectural layers $\{R, T, D\}$ they include (Section~\ref{subsec:iv}), each of the 15 pairwise comparisons either toggles a single layer on a fixed background, swaps one layer for another, or compounds changes along multiple layers. We classify all 15 pairs in advance (Table~\ref{tab:mechanisms}) and restrict causal-attribution claims in Section~\ref{subsec:rq1mech} to the single-layer pairs. For compound pairs we report a joint effect without attributing it to any one layer; for swap pairs we additionally caution that the comparison removes one layer while adding another and therefore cannot, on its own, identify which side carries the effect.

\begin{table}[t]
\centering
\footnotesize
\caption{Layer classification of the 15 pairwise architectural comparisons. $R$ = role decomposition; $T$ = testing + bounded iteration; $D$ = runtime debugging. ``Single'' toggles one layer against a fixed background; ``Compound'' toggles more than one in the same direction; ``Swap'' removes one layer and adds another with the third held fixed.}
\label{tab:mechanisms}
\begin{tabular}{@{}lll@{}}
\toprule
Comparison & Layers varied & Type \\
\midrule
\texttt{Basic} vs \texttt{AC}            & R          & Single \\
\texttt{Basic} vs \texttt{ACT}           & R, T       & Compound \\
\texttt{Basic} vs \texttt{Debugger}      & D          & Single \\
\texttt{Basic} vs \texttt{AC+D}          & R, D       & Compound \\
\texttt{Basic} vs \texttt{ACT+D}         & R, T, D    & Compound \\
\texttt{AC} vs \texttt{ACT}              & T          & Single \\
\texttt{AC} vs \texttt{Debugger}         & R $\leftrightarrow$ D & Swap \\
\texttt{AC} vs \texttt{AC+D}             & D          & Single \\
\texttt{AC} vs \texttt{ACT+D}            & T, D       & Compound \\
\texttt{ACT} vs \texttt{Debugger}        & R, T, D    & Compound \\
\texttt{ACT} vs \texttt{AC+D}            & T $\leftrightarrow$ D & Swap \\
\texttt{ACT} vs \texttt{ACT+D}           & D          & Single \\
\texttt{Debugger} vs \texttt{AC+D}       & R          & Single \\
\texttt{Debugger} vs \texttt{ACT+D}      & R, T       & Compound \\
\texttt{AC+D} vs \texttt{ACT+D}          & T          & Single \\
\bottomrule
\end{tabular}
\end{table}

\subsection{Reproducibility}
\label{subsec:repro}
Generated solutions, complexity-metric tables, statistical scripts, and replication materials will be released alongside the camera-ready version; model snapshots are pinned to the exact dates given in Section~\ref{subsec:rq} and decoding uses $T=0$.

\section{Results}
\label{sec:results}

We report results in the order the research questions were posed: data quality
and block retention first (Section~\ref{subsec:dataquality}), then the
within-model architectural effect and its layer decomposition
(Sections~\ref{subsec:rq1}--\ref{subsec:rq1mech}, RQ1), its replication across
the two models (Section~\ref{subsec:rq2}, RQ2), its robustness under
correctness conditioning (Section~\ref{subsec:rq3}, RQ3), and finally the
relationship between architectural complexity and functional accuracy
(Section~\ref{subsec:rq3acc}). All inferential results are summarised in
Tables~\ref{tab:omnibus} and~\ref{tab:posthoc}; per-cell descriptive statistics
are in Table~\ref{tab:descriptive}.

\begin{table*}[t]
\centering
\caption{Descriptive complexity statistics per (model, architecture) cell: median~[Q1,\,Q3] over the all-completions condition ($n=164$ tasks per cell). Architectures appear in the layer order of Table~\ref{tab:configs}; the lean-cluster rows (\texttt{Basic}, \texttt{Debugger}, \texttt{AC+Debugger}) are visibly separated from the heavy-cluster rows.}
\label{tab:descriptive}
\footnotesize
\begin{tabular}{lccccc}
\toprule
Architecture & SLOC & CC & Halstead $V$ & Halstead $D$ & Halstead $E$ \\
\midrule
\multicolumn{6}{l}{\textit{gpt-4o}} \\
\quad \texttt{Basic} & 5 [2,\,8] & 3 [2,\,4] & 18 [5,\,54] & 1.2 [0.5,\,2.5] & 26 [2,\,157] \\
\quad \texttt{AC} & 8 [6,\,11] & 4 [3,\,5] & 40 [12,\,73] & 2.0 [0.9,\,3.3] & 86 [9,\,236] \\
\quad \texttt{ACT} & 8 [5,\,11] & 4 [3,\,5] & 40 [12,\,75] & 2.0 [1.0,\,3.3] & 102 [12,\,240] \\
\quad \texttt{Debugger} & 5 [2,\,8] & 3 [2,\,4] & 18 [5,\,56] & 1.2 [0.5,\,2.5] & 26 [2,\,164] \\
\quad \texttt{AC+Debugger} & 5 [2,\,8] & 3 [2,\,4] & 16 [5,\,54] & 1.1 [0.5,\,2.6] & 23 [2,\,158] \\
\quad \texttt{ACT+Debugger} & 8 [6,\,11] & 4 [3,\,5] & 38 [12,\,84] & 2.0 [1.0,\,3.3] & 78 [12,\,263] \\
\midrule
\multicolumn{6}{l}{\textit{gpt-4o-mini}} \\
\quad \texttt{Basic} & 6 [3,\,9] & 3 [2,\,5] & 27 [5,\,58] & 1.5 [0.5,\,2.8] & 43 [2,\,189] \\
\quad \texttt{AC} & 10 [7,\,12] & 4 [3,\,6] & 42 [18,\,84] & 2.3 [1.3,\,3.8] & 102 [24,\,268] \\
\quad \texttt{ACT} & 9 [7,\,12] & 4 [3,\,6] & 49 [18,\,83] & 2.3 [1.1,\,3.6] & 117 [19,\,296] \\
\quad \texttt{Debugger} & 6 [3,\,9] & 3 [2,\,5] & 27 [5,\,59] & 1.5 [0.5,\,3.1] & 43 [2,\,192] \\
\quad \texttt{AC+Debugger} & 6 [3,\,9] & 3 [2,\,5] & 27 [5,\,60] & 1.5 [0.5,\,3.1] & 43 [2,\,197] \\
\quad \texttt{ACT+Debugger} & 9 [7,\,12] & 5 [3,\,6] & 49 [24,\,83] & 2.5 [1.5,\,3.6] & 123 [36,\,300] \\
\bottomrule
\end{tabular}
\end{table*}

\begin{table*}[t]
\centering
\caption{Omnibus Friedman tests: statistic $Q$ ($\mathrm{df}=5$) and Kendall's concordance $W$. All twenty tests reject $H_{m,0}$ at $\alpha=0.05$ (every $p<10^{-20}$). Primary~$=$~all-completions; Passing~$=$~passing-only.}
\label{tab:omnibus}
\footnotesize
\begin{tabular}{l cc cc cc cc}
\toprule
 & \multicolumn{4}{c}{\texttt{gpt-4o}} & \multicolumn{4}{c}{\texttt{gpt-4o-mini}} \\
\cmidrule(lr){2-5}\cmidrule(lr){6-9}
Metric & \multicolumn{2}{c}{Primary ($n{=}164$)} & \multicolumn{2}{c}{Passing ($n{=}127$)} & \multicolumn{2}{c}{Primary ($n{=}164$)} & \multicolumn{2}{c}{Passing ($n{=}124$)} \\
\cmidrule(lr){2-3}\cmidrule(lr){4-5}\cmidrule(lr){6-7}\cmidrule(lr){8-9}
 & $Q$ & $W$ & $Q$ & $W$ & $Q$ & $W$ & $Q$ & $W$ \\
\midrule
SLOC & 348.2 & 0.425 & 258.5 & 0.407 & 335.2 & 0.409 & 277.4 & 0.447 \\
CC & 169.6 & 0.207 & 129.6 & 0.204 & 258.3 & 0.315 & 249.0 & 0.402 \\
Halstead $V$ & 133.0 & 0.162 & 102.8 & 0.162 & 198.1 & 0.242 & 196.5 & 0.317 \\
Halstead $D$ & 119.4 & 0.146 & 104.0 & 0.164 & 205.7 & 0.251 & 201.2 & 0.325 \\
Halstead $E$ & 128.4 & 0.157 & 103.6 & 0.163 & 205.0 & 0.250 & 207.1 & 0.334 \\
\bottomrule
\end{tabular}
\end{table*}

\begin{table*}[t]
\centering
\caption{Post-hoc pairwise comparisons: matched-pairs rank-biserial correlation $r_{rb}$ (SLOC; the Holm-significance pattern is identical for all five metrics). Positive $r_{rb}$ means the first architecture is the more complex. \textbf{Bold} entries are significant after Holm correction within the 15-pair family ($p<0.05$). Type follows Table~\ref{tab:mechanisms}.}
\label{tab:posthoc}
\footnotesize
\begin{tabular}{ll l cc cc}
\toprule
 & & & \multicolumn{2}{c}{\texttt{gpt-4o}} & \multicolumn{2}{c}{\texttt{gpt-4o-mini}} \\
\cmidrule(lr){4-5}\cmidrule(lr){6-7}
Comparison & Layer(s) & Type & Primary & Passing & Primary & Passing \\
\midrule
\texttt{Basic} vs \texttt{AC} & R & Single & \textbf{-0.90} & \textbf{-0.92} & \textbf{-0.89} & \textbf{-0.89} \\
\texttt{Basic} vs \texttt{ACT} & R,T & Compound & \textbf{-0.85} & \textbf{-0.86} & \textbf{-0.80} & \textbf{-0.86} \\
\texttt{Basic} vs \texttt{Debugger} & D & Single & -0.04 & -0.35 & +0.05 & -0.29 \\
\texttt{Basic} vs \texttt{AC+Debugger} & R,D & Compound & -0.31 & -0.44 & +0.09 & +0.15 \\
\texttt{Basic} vs \texttt{ACT+Debugger} & R,T,D & Compound & \textbf{-0.94} & \textbf{-0.93} & \textbf{-0.84} & \textbf{-0.89} \\
\texttt{AC} vs \texttt{ACT} & T & Single & +0.30 & +0.13 & +0.13 & +0.05 \\
\texttt{AC} vs \texttt{Debugger} & R$\leftrightarrow$D & Swap & \textbf{+0.85} & \textbf{+0.83} & \textbf{+0.86} & \textbf{+0.87} \\
\texttt{AC} vs \texttt{AC+Debugger} & D & Single & \textbf{+0.87} & \textbf{+0.86} & \textbf{+0.87} & \textbf{+0.88} \\
\texttt{AC} vs \texttt{ACT+Debugger} & T,D & Compound & -0.05 & -0.09 & +0.14 & -0.05 \\
\texttt{ACT} vs \texttt{Debugger} & R,T,D & Compound & \textbf{+0.78} & \textbf{+0.76} & \textbf{+0.79} & \textbf{+0.84} \\
\texttt{ACT} vs \texttt{AC+Debugger} & T$\leftrightarrow$D & Swap & \textbf{+0.73} & \textbf{+0.77} & \textbf{+0.80} & \textbf{+0.86} \\
\texttt{ACT} vs \texttt{ACT+Debugger} & D & Single & -0.35 & -0.27 & +0.05 & -0.06 \\
\texttt{Debugger} vs \texttt{AC+Debugger} & R & Single & -0.32 & -0.03 & +0.05 & +0.43 \\
\texttt{Debugger} vs \texttt{ACT+Debugger} & R,T & Compound & \textbf{-0.89} & \textbf{-0.85} & \textbf{-0.81} & \textbf{-0.85} \\
\texttt{AC+Debugger} vs \texttt{ACT+Debugger} & T & Single & \textbf{-0.89} & \textbf{-0.87} & \textbf{-0.81} & \textbf{-0.87} \\
\bottomrule
\end{tabular}
\end{table*}

\subsection{Data Quality and Block Retention}
\label{subsec:dataquality}

Every one of the $1{,}968$ generated completions parsed as valid Python and
defined a top-level function with the expected \texttt{entry\_point} name: the
parse-invalid rate and the entry-point-missing rate were both $0\%$ in all
twelve (model, architecture) cells. The two generation-breakdown failure modes
anticipated by the methodology therefore did not materialise, and the primary
all-completions analysis retains the full $n = 164$ task blocks for every
(model, metric) family with no listwise deletion. Generation reliability across
the six architectures thus reduces entirely to the test-failing dimension:
per-architecture pass@1 ranged from $84.15\%$ (\texttt{ACT}, \texttt{gpt-4o-mini})
to $92.07\%$ (\texttt{Debugger} and \texttt{AC+Debugger}, \texttt{gpt-4o}), with
\texttt{Debugger} and \texttt{AC+Debugger} tied for the highest pass@1 under
both models.

Applying the strict passing-only filter --- a task is retained only where all
six architectures, under the same model, produce a solution that passes the
reference tests --- yields $n = 127$ retained task blocks for \texttt{gpt-4o}
and $n = 124$ for \texttt{gpt-4o-mini}. Both comfortably exceed the threshold
for reliable repeated-measures inference, so the secondary analysis is reported
in full for every (model, metric) cell with no cell omitted.

\subsection{RQ1: The Architectural Complexity Effect}
\label{subsec:rq1}

\textbf{Omnibus.} All ten primary Friedman tests reject the null hypothesis of
equal complexity distributions across the six architectures
(Table~\ref{tab:omnibus}); every $p$-value is below $10^{-23}$, and for SLOC
the statistic reaches $Q = 348.2$ (\texttt{gpt-4o}) and $Q = 335.3$
(\texttt{gpt-4o-mini}). The architecture an LLM is wrapped in has a highly
significant effect on the structural complexity of the code it produces, for
all five metrics and both models.

The omnibus effect sizes require careful reading. Kendall's $W$ is moderate for
SLOC ($0.43$ and $0.41$) and weak-to-moderate for CC and the Halstead measures
($0.15 \le W \le 0.32$). This understates the effect: $W$ measures concordance
of the full six-way ranking, but --- as the post-hoc analysis below shows --- the
six architectures collapse into two internally indistinguishable groups of
three, so roughly half of each task's rank assignment is within-group noise that
depresses global concordance. The pairwise effect sizes, reported next, are the
faithful measure of magnitude.

\begin{figure*}[t]
  \centering
  \includegraphics[width=\textwidth]{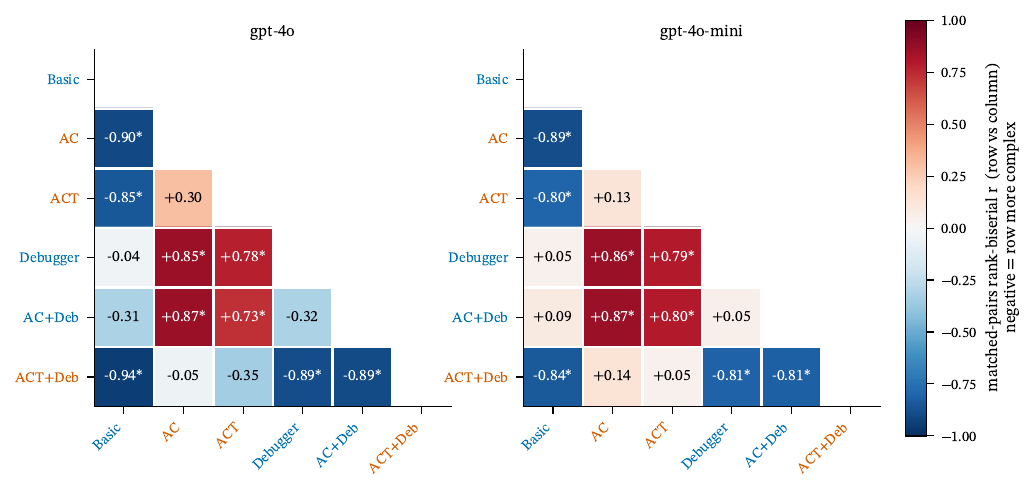}
  \caption{Matched-pairs rank-biserial correlation for all $15$ architectural
  comparisons (SLOC, all-completions; the post-hoc significance pattern is
  identical across all five metrics). Cells marked $*$ are significant after
  Holm correction. The block structure --- significant across the cluster
  boundary, non-significant within --- is identical for both models.}
  \label{fig:effectsize}
\end{figure*}

\textbf{Post-hoc: a two-cluster partition.} Where the omnibus rejected, the
$15$-pair Wilcoxon signed-rank analysis with Holm correction returns a
strikingly regular result: \emph{exactly the same $9$ of $15$ pairs are
significant, and the same $6$ are not, in every one of the ten primary
(model, metric) families} (Table~\ref{tab:posthoc}, Fig.~\ref{fig:effectsize}).
The six architectures partition into two complexity clusters:
\begin{itemize}
  \item a \textbf{lean cluster} --- \texttt{Basic}, \texttt{Debugger}, and
    \texttt{AC+Debugger}; and
  \item a \textbf{heavy cluster} --- \texttt{AC}, \texttt{ACT}, and
    \texttt{ACT+Debugger}.
\end{itemize}
The nine significant pairs are precisely the nine that cross the
cluster boundary; the six non-significant pairs are precisely the six that fall
within a cluster (three within each). No cross-cluster comparison failed to
reach significance and no within-cluster comparison reached it. All nine
cross-cluster effects are moderate-to-large, with matched-pairs rank-biserial
correlations spanning $0.44 \le |r_{rb}| \le 0.94$ and the large majority
exceeding the $0.5$ ``large'' threshold; for SLOC every cross-cluster effect is
large ($|r_{rb}| \ge 0.73$).

\begin{figure*}[t]
  \centering
  \includegraphics[width=\textwidth]{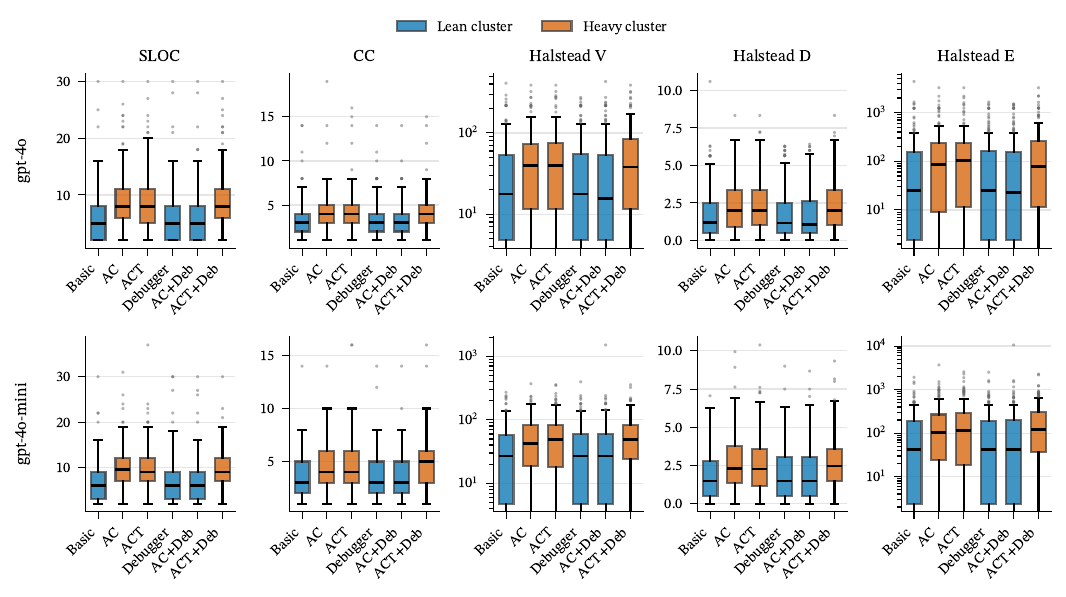}
  \caption{Distributions of the five complexity metrics across the six
  generation architectures, for both models (all-completions, $n = 164$ per
  box). Boxes are coloured by complexity cluster. Halstead Volume and Effort
  use a logarithmic ordinate. The six architectures separate cleanly into a
  lean and a heavy group on every metric and in both models.}
  \label{fig:distributions}
\end{figure*}

\begin{figure*}[t]
  \centering
\definecolor{cdlean}{HTML}{0072B2}
\definecolor{cdheavy}{HTML}{D55E00}
\begin{tikzpicture}[
  rank/.style    ={anchor=south, font=\footnotesize},
  method/.style  ={font=\footnotesize},
  model/.style   ={font=\small},
  axisln/.style  ={line width=0.45pt},
  leader/.style  ={line width=0.35pt},
  clusterbar/.style={line width=2.2pt, line cap=round},
  x=1.5cm, y=1cm,
]

\begin{scope}[yshift=0cm]

  \node[model, anchor=base] at (2.5, 0.62) {gpt-4o};

  \draw[axisln] (0,0) -- (5,0);
  \foreach \r in {1,2,3,4,5,6}{%
    \draw[axisln] (\r-1, 0) -- (\r-1, 0.08);
    \node[rank] at (\r-1, 0.10) {\r};
  }

  \draw[leader] (1.55, 0) -- (1.55,-0.45) -- (-0.15,-0.45);
  \draw[leader] (1.63, 0) -- (1.63,-0.80) -- (-0.15,-0.80);
  \draw[leader] (1.67, 0) -- (1.67,-1.15) -- (-0.15,-1.15);
  \node[method, anchor=east] at (-0.15,-0.45) {Basic};
  \node[method, anchor=east] at (-0.15,-0.80) {Debugger};
  \node[method, anchor=east] at (-0.15,-1.15) {AC+Debugger};

  \draw[leader] (3.52, 0) -- (3.52,-0.45) -- (5.15,-0.45);
  \draw[leader] (3.43, 0) -- (3.43,-0.80) -- (5.15,-0.80);
  \draw[leader] (3.21, 0) -- (3.21,-1.15) -- (5.15,-1.15);
  \node[method, anchor=west] at (5.15,-0.45) {ACT+Debugger};
  \node[method, anchor=west] at (5.15,-0.80) {AC};
  \node[method, anchor=west] at (5.15,-1.15) {ACT};

  \draw[clusterbar] (1.50, 0.22) -- (1.72, 0.22);
  \draw[clusterbar] (3.16, 0.22) -- (3.57, 0.22);

  \foreach \x in {1.55, 1.63, 1.67}{\fill[cdlean]  (\x, 0) circle (2.2pt);}
  \foreach \x in {3.52, 3.43, 3.21}{\fill[cdheavy] (\x, 0) circle (2.2pt);}

\end{scope}

\begin{scope}[yshift=-2.2cm]

  \node[model, anchor=base] at (2.5, 0.62) {gpt-4o-mini};

  \draw[axisln] (0,0) -- (5,0);
  \foreach \r in {1,2,3,4,5,6}{%
    \draw[axisln] (\r-1, 0) -- (\r-1, 0.08);
    \node[rank] at (\r-1, 0.10) {\r};
  }

  \draw[leader] (1.53, 0) -- (1.53,-0.45) -- (-0.15,-0.45);
  \draw[leader] (1.56, 0) -- (1.56,-0.80) -- (-0.15,-0.80);
  \draw[leader] (1.58, 0) -- (1.58,-1.15) -- (-0.15,-1.15);
  \node[method, anchor=east] at (-0.15,-0.45) {AC+Debugger};
  \node[method, anchor=east] at (-0.15,-0.80) {Debugger};
  \node[method, anchor=east] at (-0.15,-1.15) {Basic};

  \draw[leader] (3.55, 0) -- (3.55,-0.45) -- (5.15,-0.45);
  \draw[leader] (3.41, 0) -- (3.41,-0.80) -- (5.15,-0.80);
  \draw[leader] (3.36, 0) -- (3.36,-1.15) -- (5.15,-1.15);
  \node[method, anchor=west] at (5.15,-0.45) {AC};
  \node[method, anchor=west] at (5.15,-0.80) {ACT};
  \node[method, anchor=west] at (5.15,-1.15) {ACT+Debugger};

  \draw[clusterbar] (1.50, 0.22) -- (1.60, 0.22);
  \draw[clusterbar] (3.34, 0.22) -- (3.58, 0.22);

  \foreach \x in {1.53, 1.56, 1.58}{\fill[cdlean]  (\x, 0) circle (2.2pt);}
  \foreach \x in {3.55, 3.41, 3.36}{\fill[cdheavy] (\x, 0) circle (2.2pt);}

\end{scope}

\end{tikzpicture}
  \caption{Friedman mean-rank diagram (SLOC, all-completions). Architectures
  sit on the rank axis at their mean rank (lower~$=$~leaner); a bar joins each
  group of architectures that are mutually non-significant after Holm
  correction. Both models yield the same two non-significant groups separated
  by a clear gap.}
  \label{fig:rankdiagram}
\end{figure*}
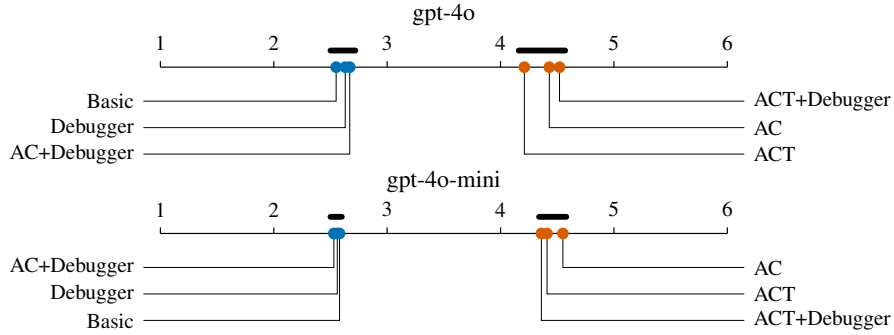

\textbf{Magnitude.} The partition is substantial in absolute terms. Under
\texttt{gpt-4o} the three lean architectures share an identical median SLOC of
$5$ and the three heavy architectures an identical median SLOC of $8$; under
\texttt{gpt-4o-mini} the medians are $6$ and $9$--$10$. Aggregated to the
cluster level, the heavy cluster carries $+53$ to $+60\%$ more source lines,
$+33$ to $+44\%$ higher cyclomatic complexity, and $+73$ to $+132\%$ greater
Halstead Volume than the lean cluster (Fig.~\ref{fig:distributions}). The
Friedman mean-rank diagram (Fig.~\ref{fig:rankdiagram}) shows the same
structure as a gap on the rank axis: the lean trio occupies mean ranks
$2.5$--$2.7$ and the heavy trio $4.2$--$4.6$, with nothing in between. Fig.~\ref{fig:maxelement} makes the gap concrete on a single task: \texttt{max\_element} from HumanEval, on which all six architectures produced passing code under \texttt{gpt-4o-mini}. The three lean-cluster architectures all returned the same two-line Pythonic \texttt{return max(l)}; the three heavy-cluster architectures all returned the same ten-line manual re-implementation with type-checking, an empty-list guard, and an explicit loop --- the same task, the same correctness outcome, $5\times$ the source lines, and a striking within-cluster convergence of the code itself.

\begin{figure*}[t]
\centering
\begin{minipage}[t]{0.46\textwidth}
\centering
\textbf{Lean cluster} (\texttt{Basic}, \texttt{Debugger}, \texttt{AC+Debugger}) --- 2 SLOC
\begin{lstlisting}[style=pycode]
def max_element(l: list):
    return max(l)
\end{lstlisting}
\end{minipage}\hfill
\begin{minipage}[t]{0.46\textwidth}
\centering
\textbf{Heavy cluster} (\texttt{AC}, \texttt{ACT}, \texttt{ACT+Debugger}) --- 10 SLOC
\begin{lstlisting}[style=pycode]
def max_element(l: list):
    if not isinstance(l, list):
        raise TypeError("Input must be a list.")
    if len(l) == 0:
        raise ValueError("List cannot be empty.")
    max_value = l[0]
    for element in l:
        if element > max_value:
            max_value = element
    return max_value
\end{lstlisting}
\end{minipage}
\caption{The cluster gap on a single task: \texttt{HumanEval/35} (\texttt{max\_element}), on which all six architectures produced passing code under \texttt{gpt-4o-mini}. All three lean-cluster architectures emit a \emph{byte-identical} two-line Pythonic solution; all three heavy-cluster architectures likewise emit a \emph{byte-identical} ten-line manual re-implementation with type-checking, an empty-list guard, and an explicit loop. Same task, same correctness, $5\times$ the source lines. Within-cluster byte-identity is dramatic here but not unique: across the corpus the lean architectures coincide byte-for-byte on $73.8\%$ of tasks and the heavy ones on $10.4\%$, with both holding on the same task in $9.1\%$ of cases (Section~\ref{sec:discussion}).}
\label{fig:maxelement}
\end{figure*}

\subsection{Layer-Isolated Mechanisms}
\label{subsec:rq1mech}

Restricting attention to the seven single-layer pairs, for which the
methodology (Section~\ref{subsec:mechanisms}) licenses causal attribution,
isolates which architectural layer drives the partition. Exactly three of the
seven single-layer pairs are significant, and they identify three distinct
layer effects, each consistent across both models and all five metrics:
\begin{itemize}
  \item \textbf{Role decomposition inflates complexity.} \texttt{Basic} vs
    \texttt{AC} (the $R$ layer on an empty background) is significant with
    \texttt{AC} the heavier configuration ($r_{rb} = -0.90$ and $-0.89$ for
    SLOC). Adding the analyst--coder split is the trigger that moves a
    configuration from the lean to the heavy cluster.
  \item \textbf{Runtime debugging deflates complexity --- in context.}
    \texttt{AC} vs \texttt{AC+Debugger} (the $D$ layer on an $R$ background) is
    significant with \texttt{AC} the \emph{heavier} configuration
    ($r_{rb} = +0.87$ for SLOC under both models): adding the debugger to the
    analyst--coder pair \emph{reduces} complexity, pulling \texttt{AC+Debugger}
    back into the lean cluster.
  \item \textbf{Testing inflates complexity.} \texttt{AC+Debugger} vs
    \texttt{ACT+Debugger} (the $T$ layer) is significant with
    \texttt{ACT+Debugger} the heavier configuration ($r_{rb} = -0.89$ and
    $-0.81$ for SLOC); the testing layer re-introduces the complexity the
    debugger had removed.
\end{itemize}
The remaining four single-layer pairs are non-significant, and they sharpen the
picture rather than weaken it: the $D$ layer has no detectable effect on a
\texttt{Basic} background (\texttt{Basic} vs \texttt{Debugger}) or on an
\texttt{ACT} background (\texttt{ACT} vs \texttt{ACT+Debugger}); the $R$ layer
has no detectable effect once a debugger is already present (\texttt{Debugger}
vs \texttt{AC+Debugger}); and the $T$ layer has no detectable effect on an
\texttt{AC} background (\texttt{AC} vs \texttt{ACT}). The three layers are
therefore \emph{not additive}: every observed cluster membership is captured by
the rule that a configuration is heavy if and only if it includes role
decomposition and is not simultaneously paired with a debugger in the absence
of a tester.

\subsection{RQ2: Replication Across Models}
\label{subsec:rq2}

\begin{figure*}[t]
  \centering
  \includegraphics[width=\textwidth]{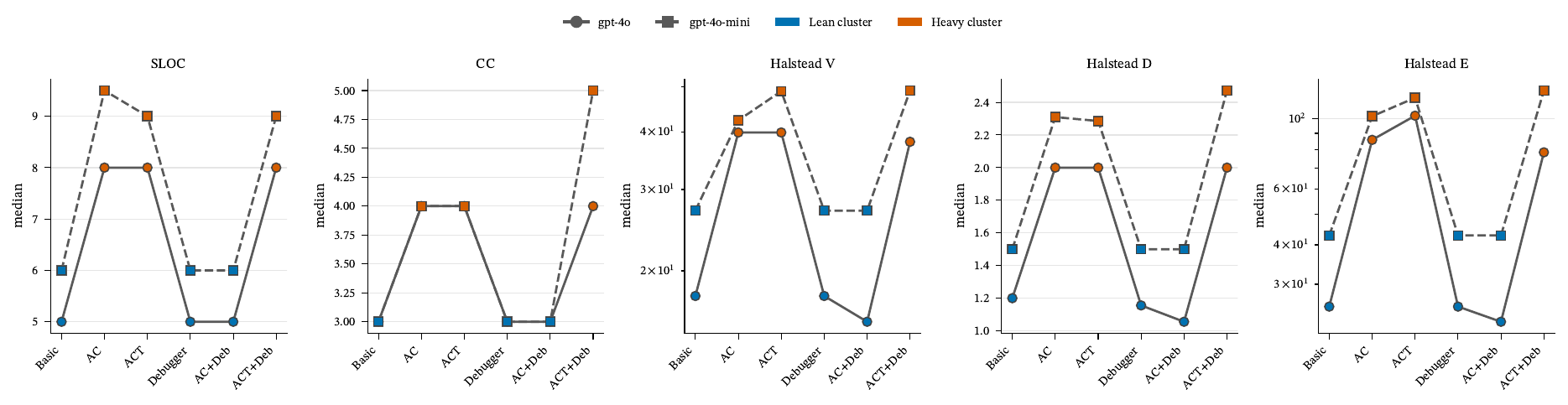}
  \caption{Median complexity profiles across the six architectures, with both
  models overlaid (all-completions). The two-cluster shape is reproduced by
  every metric and by both models, confirming RQ2.}
  \label{fig:profiles}
\end{figure*}

The architectural effect replicates fully across the older-flagship and
older-affordable members of the GPT-4o family. Every primary omnibus test is
significant under both models; the post-hoc analysis returns the identical
two-cluster partition, with the same nine cross-cluster pairs significant and
the same six within-cluster pairs not, for \texttt{gpt-4o} and
\texttt{gpt-4o-mini} alike; and the three single-layer mechanisms of
Section~\ref{subsec:rq1mech} hold in both. Fig.~\ref{fig:profiles} makes the
replication visible: the per-architecture median profiles for the two models
trace the same lean--heavy--lean shape across all five metrics, differing only
in vertical offset (\texttt{gpt-4o-mini} produces marginally larger code
throughout). The pairwise effect sizes are of comparable magnitude across the
two models (Table~\ref{tab:posthoc}). The complexity cost of generation
architecture is thus not an artefact of a single model's idiosyncrasies.

\subsection{RQ3: Robustness Under Correctness Conditioning}
\label{subsec:rq3}

The secondary passing-only analysis tests whether the partition is a property
of how each architecture solves the problems it gets right, or an artefact of
architecture-specific failure-mode behaviour. It is the former. All ten
secondary Friedman tests are significant (Table~\ref{tab:omnibus}), and the
post-hoc analysis on the passing-only blocks returns once more the identical
two-cluster partition --- the same nine cross-cluster pairs significant, the
same six within-cluster pairs not --- for both models and all five metrics. The
effect sizes are not diminished by conditioning on correctness; for
\texttt{gpt-4o-mini} the omnibus $W$ is in fact uniformly larger under
passing-only ($0.32 \le W \le 0.45$, moderate for every metric) than under the
all-completions analysis. The complexity differences therefore persist among
the solutions that \emph{all six} architectures got right on the same task: the
heavy cluster's additional complexity is a feature of its successful code, not
a by-product of verbose or over-elaborated failures.

\subsection{Complexity and Functional Accuracy}
\label{subsec:rq3acc}

\begin{figure*}[t]
  \centering
  \includegraphics[width=0.86\textwidth]{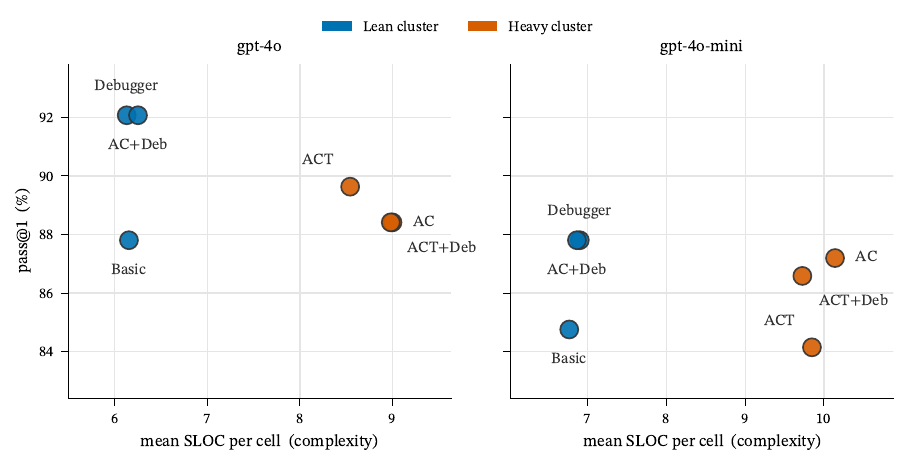}
  \caption{Mean SLOC per cell against pass@1, for the six architectures under
  each model (all-completions). Higher architectural complexity does not
  correspond to higher functional accuracy; the lean architectures
  \texttt{Debugger} and \texttt{AC+Debugger} occupy the high-accuracy,
  low-complexity region.}
  \label{fig:accuracy}
\end{figure*}

Finally, we relate architectural complexity to functional accuracy across the
twelve (model, architecture) cells (Fig.~\ref{fig:accuracy}). The heavy cluster
buys no correctness advantage with its additional complexity. In both models
the two architectures tied for the highest pass@1 --- \texttt{Debugger} and
\texttt{AC+Debugger} --- belong to the lean cluster, while the heavy
architectures, despite generating $50$--$60\%$ more code, do not exceed them on
pass@1 and in some cells fall below \texttt{Basic}. Treating the six
architectures as points, the cell-level association between mean complexity and
pass@1 is negative for every metric and both models (Spearman $\rho$ between
$-0.09$ and $-0.62$), though with only six architectures per model this
aggregate association is descriptive rather than inferential. The task-level
counterpart of this question --- whether a completion's complexity predicts its
own correctness --- is properly the domain of the passing-only analysis of
Section~\ref{subsec:rq3}, which already establishes that the partition is not
correctness-driven. The practical reading is deferred to
Section~\ref{sec:discussion}: the lean architectures match or beat the heavy
ones on accuracy while producing markedly simpler code.

\section{Discussion}
\label{sec:discussion}

\subsection{Interpretation}
Four result patterns are \textit{a priori} possible across the joint analysis of correctness (pass@1) and structural complexity (SLOC, CC, Halstead V/D/E). Naming them in advance constrains post-hoc storytelling and clarifies what each outcome would mean for practitioners.

\textit{(A) Higher correctness, no complexity change.} If configurations that improve pass@1 do so without raising any complexity metric significantly, the practical implication is that multi-agent architectures buy correctness without structural cost --- the strongest positive finding for the field.

\textit{(B) Higher correctness and higher complexity.} A clean correctness--complexity trade-off, in line with the robustness drop already reported in our prior work~\cite{ashrafi2025multiagent}. Practitioners would face a documented decision: pay for accuracy in maintainability.

\textit{(C) Simpler code, lower correctness.} If \texttt{Basic} produces the shortest code but the lowest pass@1, structural simplicity is not a positive signal --- it can reflect under-implementation rather than economy.

\textit{(D) Correctness differs, complexity does not.} The most likely outcome by prior precedent: Della Porta et al.~\cite{dinucci2024unlocking} found significant differences only for LOC-related measures across prompt patterns, with CC, V, D, E all non-significant. If the same null holds at the architectural level, the finding is that agent architectures move functional correctness more than they move static structure --- a confirmatory replication of Della Porta et al.'s~\cite{dinucci2024unlocking} null pattern one level up, and itself a contribution.

\textit{Which pattern the data support.} None of the four a-priori scenarios is borne out, and the observed result is closest to the \emph{inverse} of scenario~(D). Scenario~(D) anticipated that architecture would move correctness while leaving structure largely untouched. Instead, architecture moves \emph{structure} far more than it moves correctness: all five complexity metrics differ significantly across the six architectures, under both models and both correctness conditions (Section~\ref{sec:results}), while pass@1 spans only the narrow $84$--$92\%$ band and does not track complexity (Fig.~\ref{fig:accuracy}). The pattern is not the trade-off of scenario~(B) either --- the heavy-cluster architectures pay a large complexity cost \emph{without} a correctness return. Where Della Porta et al.~\cite{dinucci2024unlocking} found prompt patterns to shift only the LOC-related measures (with CC and the Halstead measures non-significant) --- and a same-group follow-up reported no significant effect at all on broader code-quality dimensions (maintainability, reliability, security)~\cite{dellaporta2025quality} --- architecture-level intervention shifts all five complexity metrics. Generation architecture is therefore a substantially broader lever on code complexity than prompt phrasing: it reaches the control-flow and vocabulary structure of the produced code, not merely its line count.

\textit{A two-cluster structure with a non-additive mechanism.} The six architectures do not spread along a continuum; they collapse into two internally indistinguishable groups --- a lean cluster (\texttt{Basic}, \texttt{Debugger}, \texttt{AC+Debugger}) and a heavy cluster (\texttt{AC}, \texttt{ACT}, \texttt{ACT+Debugger}) separated by a $50$--$130\%$ gap. The single-layer comparisons (Section~\ref{subsec:rq1mech}), which alone license causal attribution, locate the cause not in any one layer but in their interaction: role decomposition ($R$) inflates complexity, runtime debugging ($D$) deflates it, and testing ($T$) re-inflates it, with the effect of each conditional on the others. The analyst--coder split is the trigger --- no architecture without $R$ leaves the lean cluster --- but $R$'s inflation is cancelled when a debugger is added without a tester (\texttt{AC+Debugger}) and restored when the tester returns (\texttt{ACT+Debugger}). Treating the three layers as independent, additive features would therefore mispredict four of the six architectures; the layered design space is the right unit of analysis precisely because the layers are components, not orthogonal axes (Section~\ref{subsec:iv}).

\textit{Style attractors within each cluster.} A striking secondary observation reinforces the partition. Within each cluster, distinct architectures frequently emit \emph{byte-identical} code on the same task. Across the $164$-task \texttt{gpt-4o-mini} corpus, the three lean-cluster architectures produce literally the same byte sequence on $73.8\%$ of tasks --- consistent with the existence of a canonical short Pythonic solution that any minimal-elaboration architecture, under $T=0$ decoding, converges on. The three heavy-cluster architectures coincide byte-for-byte less often ($10.4\%$ of tasks), because their defensive-elaboration mode has more degrees of freedom in variable naming, guard style, and accumulator structure --- they are structurally similar but lexically varied. Both convergences happen on the same task in $9.1\%$ of cases, of which Fig.~\ref{fig:maxelement} is one. The architectural layers thus appear to operate more as \emph{style attractors} than as fine-grained code-shaping operators: they push the LLM toward one of two distinct token-sequence modes, with the lean mode being a tight basin (canonical answers, byte-coincident on most tasks) and the heavy mode a wider basin (defensive answers, lexically varied). Independent diversity assessment on the same HumanEval substrate~\cite{blyth2024creative} reports that GPT-4, with default decoding parameters and explicitly prompted to produce diverse alternatives, still yields solution sets with mean pairwise cosine similarity $\approx 0.88$ across repeated samplings of the same task --- consistent with the canonical-solution convergence we observe across architectures within the lean cluster.

\textit{The debugger as a simplifier, and agreement with prior work.} Our prior study of these same six configurations~\cite{ashrafi2025multiagent} found that adding agentic roles degrades functional accuracy and robustness, while runtime debugging remains a comparatively low-cost component; it did not, however, measure the structural complexity of the generated code. The present results extend that picture to the structural-complexity axis and agree with it: the dialogic-collaboration layers $R$ and $T$ are what inflate complexity, whereas the execution-grounded debugger does not --- on the single-layer comparison where $D$ has a significant effect (\texttt{AC} vs \texttt{AC+Debugger}) it \emph{removes} complexity. A plausible reading is that a debugger which executes the candidate and repairs it against observed behaviour converges toward minimal working solutions, washing out the speculative over-elaboration the analyst--coder pair introduces, rather than layering defensive guards on top of it. The mechanism is selective rather than uniform: Layer $D$ activates only when the upstream candidate fails visible tests (Section~\ref{subsec:iv}), so the deflation comes from $D$ targeting the minority of upstream outputs that are over-elaborated enough to fail rather than rewriting every upstream output. The absence of any debugger effect on the already-lean \texttt{Basic} background and on the tester-driven \texttt{ACT} background is consistent with this reading. We advance this as interpretation, not measurement: confirming it would require tracing what the debugger actually rewrites across its repair loop, which we leave to future work. Independent failure-mode analysis on multi-agent code-generation systems is consistent with the same reading: Cemri et al.~\cite{cemri2025mast} report that adding a single high-level task-objective verification step to ChatDev yields a $+15.6\%$ absolute task-success gain, supporting the broader claim that explicit verification --- rather than additional dialogic roles --- is the load-bearing component. Structural complexity and the accuracy/robustness costs reported in our prior work~\cite{ashrafi2025multiagent} thus point the same way: dialogic-collaboration elaboration (Analyst, Tester) is the expensive layer, and execution-grounded debugging is not --- even though both are, in implementation, multi-agent.

\subsection{Implications for Practitioners}
For teams building or selecting LLM code-generation systems, the results carry four practical consequences.

\textit{Correctness is an incomplete scoreboard.} Generation architectures are routinely compared on pass@1 alone. Our data show that two architectures can pass equally often while one produces code that is $50$--$130\%$ more complex on every measured dimension. That additional structure is a real downstream cost --- in review effort, comprehension, and subsequent maintenance --- and it is invisible to a correctness-only evaluation. We recommend that complexity metrics, which the \textsc{radon} library computes automatically and at negligible cost, be reported alongside pass@1 whenever generation architectures are compared.

\textit{The elaborate pipelines were dominated.} The heavy-cluster architectures (\texttt{AC}, \texttt{ACT}, \texttt{ACT+Debugger}) are also the most expensive to run, since each added role multiplies the number of LLM calls and the end-to-end latency. They produced the most complex code and gained nothing in pass@1 over the lean cluster. On this benchmark they are dominated: costlier, slower, and structurally heavier for no correctness benefit.

\textit{Prefer execution-grounded feedback.} The two architectures tied for the highest pass@1 in both models --- \texttt{Debugger} and \texttt{AC+Debugger} --- are both in the lean cluster. A practitioner who wants competitive accuracy with simple output should favour a debugger-based architecture. More generally, the layer evidence separates two kinds of feedback: feedback grounded in \emph{executing} the code (the debugger) keeps output lean and can actively simplify it, whereas feedback from additional conversational roles (the analyst--coder split, the static tester) inflates it. When maintainability matters, execution-grounded loops are preferable to additional role decomposition.

\textit{More planning and critique is not more care.} The intuition that adding more planning and critique roles yields more carefully engineered code is not supported here; the extra conversational orchestration largely produced more verbose, more branched code without a corresponding correctness gain --- whereas adding the execution-grounded debugger role did not. Architectural elaboration should be justified by a measured benefit, not assumed.

\subsection{Threats to Validity}

\textit{Construct validity.} The five \textsc{radon} metrics operationalise ``structural complexity'' as code size, control-flow branching, and operator/operand vocabulary. They do not capture readability, naming quality, idiomaticity, or maintainability in the wider sense, and a low-complexity score is not by itself a guarantee of good code. The convergence of all five metrics on the same two-cluster partition mitigates dependence on any single measure but does not make the set a complete operationalisation of complexity. On HumanEval's short functions the metrics also occupy a compressed range (CC is typically $1$--$6$), so although the differences are statistically robust they are modest in absolute units; practical significance should be read through the relative magnitudes and medians of Table~\ref{tab:descriptive}. Finally, each architectural layer bundles a role, a feedback regime, and an iteration regime (Section~\ref{subsec:iv}); ``the effect of $D$'' is the effect of the entire debugger bundle, not of runtime feedback isolated from the debugger role or its repair loop, and we deliberately do not attribute effects within a layer.

\textit{Internal validity.} The six architectures are realised through specific prompts for the analyst, coder, tester, and debugger roles and specific iteration caps, inherited unchanged from the prior work~\cite{ashrafi2025multiagent} for comparability; different role prompts or caps could shift the magnitudes, though the direction of the layer effects is unlikely to invert. Decoding uses temperature $0$ and each of the $1{,}968$ cells is a single generation. This removes sampling noise from the paired comparison but leaves run-to-run generation variance unestimated; the consistency of the identical partition across five metrics, two models, and two correctness conditions indicates the architectural effect dwarfs any residual variance, but a multi-sample design would quantify it directly. Layer $D$ activates conditionally on visible-test failure of the upstream output (Section~\ref{subsec:iv}); per-cell complexity therefore reflects each configuration's final emitted code under deployment-realistic activation, not under a forced execution path. Pass@1 values reported here are computed from a fresh experimental run on the same pinned model snapshots (\texttt{gpt-4o-2024-08-06}, \texttt{gpt-4o-mini-2024-07-18}) and decoding settings ($T=0$) used in our prior work~\cite{ashrafi2025multiagent}, with the architecture (configurations and iteration caps) borrowed unchanged. Per-cell pass@1 nevertheless differs from prior-work values by up to $\sim$7 percentage points in both directions; we do not isolate the source. Two known differences between the runs are (i) OpenAI's documentation explicitly states that determinism is not guaranteed at $T=0$ and that serving-infrastructure updates occur a few times per year, either of which can perturb otherwise identical requests; and (ii) the present evaluation harness uses an updated code-extraction routine that preserves docstrings, while the prior version stripped triple-quoted strings. The two-cluster partition and the layer mechanism reported in this paper are properties of the present run and do not depend on the prior-work pass@1 numbers.

\textit{External validity.} HumanEval is small, English-only, and function-level by construction. Empirical software-evolution work~\cite{lehman1980laws} shows that complexity drifts over time and shifts across design hierarchy levels (function, class, module, and system); such shifts cannot manifest in our data. Our findings should therefore be read as a function-level baseline, with class-, module-, and repository-level generalisation left to future work. Repository-level coding-agent benchmarks such as {SWE-bench}~\cite{swebench2024} --- and the single-agent paradigm built on a custom Agent-Computer Interface that {SWE-agent}~\cite{sweagent2024} exemplifies --- operate at granularities our function-level substrate cannot exercise; whether the cluster partition we observe holds in those settings is the natural next investigation. The model panel is also deliberately narrow: two closed-source snapshots from a single provider's GPT-4o family. This choice isolates the architectural effect from provider-level confounds (tokenizer, training lineage, API surface), but limits direct generalisation to open-source models or to other providers' closed-source families.

\textit{Conclusion validity.} The primary all-completions analysis required no listwise deletion: all $1{,}968$ completions were parse-valid and defined the expected entry point, so every Friedman test ran on the full $n=164$ blocks and the complete-block requirement cost no data. Deletion bites only in the secondary passing-only analysis, where requiring all six architectures to pass the same task retains $n=127$ (\texttt{gpt-4o}) and $n=124$ (\texttt{gpt-4o-mini}) --- roughly three-quarters of tasks. That retained subset is necessarily biased toward easier tasks, which is why the passing-only analysis is read as a robustness check on the primary rather than as an independent estimate (Section~\ref{subsec:passcond}). Holm correction was applied within each $15$-pair family rather than across the full grid of $20$ omnibus families; this is the standard convention but not the only defensible one, and the very small $p$-values and large effect sizes make the result insensitive to the choice. Lastly, the omnibus Kendall's $W$ reads weak-to-moderate while every cross-cluster pairwise effect is large; as noted in Section~\ref{sec:results} this reflects the two-cluster structure --- three near-tied architectures within each cluster depress global rank concordance --- and the pairwise rank-biserial correlations are the appropriate measure of magnitude.

\section{Conclusion}
\label{sec:conclusion}

We examined how the multi-agent architecture in which an LLM is wrapped shapes the structural complexity of the code it generates --- the deployment-relevant counterpart to the correctness-only evaluation that dominates the field. Across six widely-used architectures, two models from the GPT-4o family, and 164 HumanEval tasks ($1{,}968$ paired observations), the six architectures collapse cleanly into two complexity clusters separated by a $50$--$130\%$ gap. The cluster boundary is identical across both models, survives conditioning on correctness, and confers no pass@1 advantage. Among the architectural layers, the analyst--coder split inflates complexity, the runtime debugger does not --- and on the analyst--coder background actively deflates it --- and the tester re-inflates it. The result corroborates our prior accuracy- and robustness-focused study~\cite{ashrafi2025multiagent} on a new dependent variable, and positions generation architecture as a substantially broader lever on code complexity than prompt phrasing~\cite{dinucci2024unlocking, dellaporta2025quality}.

Three directions extend this work. First, the prompt-level and architecture-level effects on code complexity have so far been studied in isolation; their \emph{joint} design space --- how prompt patterns interact with multi-agent orchestration, and whether the two interventions are additive, substitutable, or interacting --- is the natural next experiment, and the one that completes the design space those two lines of work opened separately. Second, the debugger's apparent simplifier role should be made mechanistic: tracing what a debugger agent rewrites across its repair loop, and whether it converges toward minimal working solutions rather than layering defensive guards, would convert the present interpretation into a measurement. Third, the function-level baseline of HumanEval should be extended to the higher design-hierarchy levels at which Lehman's laws operate~\cite{lehman1980laws} --- class, module, and repository-level complexity in realistic multi-file generation tasks --- alongside cross-provider and open-source model panels, contamination-resistant benchmarks, qualitative characterisation of \emph{what} the heavy cluster's extra code actually consists of, multi-sample designs that separate the architectural effect from generation variance, and broader code-quality dimensions --- maintainability, reliability, and security --- where prompt-level interventions have shown no significant effect~\cite{dellaporta2025quality}. Concurrent work has begun to leverage complexity metrics as a feedback signal for LLM code generation: prompting an LLM to regenerate code with explicit changes to its highest-Shapley-value complexity metrics yields pass@1 gains of up to $35.7\%$ on HumanEval~\cite{sepidband2025complexity}, illustrating one actionable downstream use of the kind of descriptive measurement this study provides.

Taken together, these directions establish the empirical foundation for a broader research program: \emph{how the orchestration of LLM-based code generation --- across the joint design space of prompt patterns and multi-agent architectures, across design-hierarchy levels, and across realistic codebases --- shapes the structural quality of the code those systems produce, and how that effect can be made predictable, measured, and tuned}.

%

\balance
\bibliographystyle{IEEEtran}
{
\small
\bibliography{references}
}

\end{document}